\documentclass[hyper,aps,eqsecnum,superscriptaddress,12pt]{revtex4-1}
\usepackage{bbm}
\usepackage{mathrsfs}
\usepackage{amsmath,amssymb,bm,comment}
\usepackage{hyperref}
\usepackage{graphicx}
\usepackage{bm}
\usepackage{stmaryrd}
\usepackage{color}
\usepackage{fancyhdr}

\begin{document}

\title{  Self-consistent constraints on the  collision terms in quantum kinetic theory }

\author{Shi-Yuan Wu}
\email{wushiyuan@mail.sdu.edu.cn}
\affiliation{School of Space  Sciences and Physics, Shandong University, Weihai, Shandong 264209, China}

\author{Jian-Hua Gao}
\email{gaojh@sdu.edu.cn}
\affiliation{School of Space  Sciences and Physics, Shandong University, Weihai, Shandong 264209, China}

\begin{abstract}
We derive self-consistent constraint conditions for collision terms in quantum kinetic theory using the Wigner function formalism. We present specific solutions for these collision terms that align with the constraints.  we develop  quantum kinetic theory at relaxation-time approximation. With this framework, we discuss spin polarization and electric charge separation effects in relativistic heavy-ion collisions.
\end{abstract}

\maketitle
\thispagestyle{fancy}
\renewcommand{\headrulewidth}{0pt}


\section{Introduction}

In 1872, Boltzmann, the founder of the kinetic theory,  derived  his famous  classical kinetic equation from the physical point of view \cite{Boltzmann:1872}.
The quantum analog of the Boltzmann equation was suggested by Uehling and Uhlenbeck in 1933 by purely phenomenological consideration\cite{Uehling:1933}.
It was not until 1946 that  Bogolyubov, Born, Green, Kirkwood Yvon and others   derived the  kinetic equations  within the mathematical
formalism from the Liouville equation \cite{Bogolyubov:1946}.
The quantum kinetic equations  based on  quantum field theory was derived by Martin and Schwinger \cite{Martin:1959jp,Schwinger:1960qe},
Kadanoff and Baym \cite{Baym:1961zz}, Keldysh \cite{Keldysh:1964ud} with the method of  non-equilibrium Green functions.
In the 1980s and 1990s , the gauge invariant quantum kinetic theories based on  QED or QCD  were derived
\cite{Heinz:1983nx,Heinz:1985vf,Elze:1986qd,Elze:1986hq,Vasak:1987um,Zhuang:1995pd,Ochs:1998qj,Zhuang:1998bqx,Wang:2001dm,Wang:2002qe}
in order to describe non-equilibrium and quantum effects in upcoming relativistic heavy-ion collisions at that time.
 In the 2000s, the pioneering work \cite{Liang:2004ph,Liang:2004xn,Kharzeev:2007jp,Fukushima:2008xe,Kharzeev:2007tn,Gao:2007bc,Becattini:2007sr}
 initiated the research on the chiral and spin effects in relativistic heavy-ion collisions, which
 lead to further development on the derivation of  quntum kinetic theory with chiral or spin  degree of freedom
\cite{Gao:2012ix,Stephanov:2012ki,Son:2012zy,Chen:2012ca,Manuel:2013zaa,Manuel:2014dza,Chen:2014cla,Chen:2015gta,Hidaka:2016yjf,
Mueller:2017lzw,Huang:2018wdl,Gao:2018wmr,Liu:2018xip,Gao:2019znl,Weickgenannt:2019dks,Hattori:2019ahi,Wang:2019moi,Lin:2019fqo,Li:2019qkf,Sheng:2020oqs,Guo:2020zpa,
Weickgenannt:2020aaf,Hayata:2020sqz,Weickgenannt:2021cuo,Sheng:2021kfc,Fauth:2021nwe,Luo:2021uog,ChenShiLe:2021guf,Lin:2021mvw,Chen:2021azy,Sheng:2022ssd,Fang:2022ttm,Wagner:2022amr,
Ma:2022ins,Wagner:2023cct}.

Collision terms in kinetic equation play a central role in describing the evolution in  non-equilibrium systume, and can  be derived from the BBGKY equation or  hierarchy\cite{Bogolyubov:1946} by introducing certain additional assumptions. As we all know, in general, the collision terms should satisfy the principle of
detailed balance and the constraints of charge conservation or energy-momentum conservation. In this paper, we will demonstrate that the collision terms in quantum kinetic theory
 must satisfy some extra constraints totally determined by the self-consistent nature of the quantum kinetic equation.
 With this self-consistent constraints, we will give  non-trivial specific solutions  for  the collision terms in quantum kinetic theory.
 We put the specific  solution  in the  form of   relaxation-time approximation and use it to discuss spin polarization and electric charge separation in relativistic heavy-ion collisions.

We will use the convention for the metric $g_{\mu\nu} = $ diag$(1, -1, -1, -1)$, Levi Civita tensor $\epsilon^{0123} = 1$. We choose natural units such that $\hbar = c = 1$
unless otherwise stated.

\section{Wigner function formalism}
\label{sec:wigner}
The quantum kinetic theory can be built with different formalisms.
In this work, we will adopt the Wigner function formalism \cite{Heinz:1983nx,Heinz:1985vf,Elze:1986qd,Elze:1986hq,Vasak:1987um,Zhuang:1995pd,Ochs:1998qj,Zhuang:1998bqx,Wang:2001dm,Wang:2002qe}
and restrict ourselves  to the system controlled by quantum  electrodynamics.
In quantum electrodynamics, we  define  gauge invariant density operator \cite{Elze:1986qd} as the following,
\begin{eqnarray}
\label{density}
\rho_{ab} \left(x,y\right) = \bar\psi_b(x)e^{\frac{y}{2}\cdot D^\dagger} e^{-\frac{y}{2}\cdot D}\psi_a(x),
\end{eqnarray}
where $\psi_a$ and $\bar\psi_b$ represent the electron's spinor field with the spinor indices $a$ and $b$  running from 1 to 4. The covariant derivative $D$ or  conjugate $D^\dagger$ in the covariant translation operator  are given by
\begin{eqnarray}
D_\mu=\partial_\mu^x +i A_\mu(x),\ \ \ D_\mu^\dagger=\partial^{x\dagger}_\mu - i A_\mu(x),
\end{eqnarray}
where we have absorbed the electric charge $e$ into the gauge potential $A_\mu$, and the derivatives,  $\partial_\mu^x $ and $\partial^{x\dagger}_\mu $,  denote  acting on  the right and the left
with respect to the coordinate $x$ , respectively.
We note that the density operator $\rho$ in Hilbert space is valued as a  $4\times4$ matrix in spinor space.

The Wigner function is  defined as the  Fourier transformation of the ensemble averaging of the  density operator by
\begin{eqnarray}
\label{wigner}
W(x,p)=\int\frac{d^4 y}{(2\pi)^4} e^{-ip\cdot y}  \langle  \rho \left(x,y\right) \rangle,
\end{eqnarray}
where the brackets denote the ensemble average. After ensemble average, the Wigner function
is only a $4\times4$ matrix in spinor space.  The Wigner equation satisfied by $W(x,p)$ can be derived from the Dirac equation \cite{Elze:1986qd}  and put in the following form
\begin{eqnarray}
\label{W-eq-full}
\left[m-\gamma_\mu\left(p^\mu +\frac{1}{2}i\partial^\mu_x\right)\right] W(x,p)
&=&\gamma_\mu \left[\frac{1}{2} i {C}^\mu(x,p) +  \Delta{C}^\mu(x,p) \right],
\end{eqnarray}
where  $C^\mu(x,p)$ and $\Delta C^\mu(x,p)$ are both four-vectors in Minkowski space but also $4\times4$ matrices in spinor space with  the element   defined as
\begin{eqnarray}
\label{C-full}
{ C}^\mu_{ab} &=& j_0\left(\frac{\Delta}{2}\right) \partial_\nu^p\int \frac{d^4y}{(2\pi)^4}e^{-ip\cdot y}
\langle  \bar\psi_b(x)e^{\frac{y}{2}\cdot D^\dagger} F^{\nu\mu}\left(x\right) e^{-\frac{y}{2}\cdot D}\psi_a(x)\rangle,\\
\label{DC-full}
\Delta {C}^\mu_{ab} &=&\frac{1}{2} j_1\left(\frac{\Delta}{2}\right)  \partial_\nu^p\int \frac{d^4y}{(2\pi)^4}e^{-ip\cdot y}
\langle  \bar\psi_b(x)e^{\frac{y}{2}\cdot D^\dagger} F^{\nu\mu}\left(x\right) e^{-\frac{y}{2}\cdot D}\psi_a(x)\rangle,
\end{eqnarray}
where  $j_0$ and $j_1$ are zeroth- and first-order spherical Bessel functions, respectively, and the triangle operator $\Delta\equiv \partial_p \cdot \partial_x$
denotes the mixed derivative.  The electromagnetic field tensor $ F_{\nu\mu}\left(x\right)$  is defined as usual
\begin{eqnarray}
\label{Fmunu}
 F_{\mu\nu}(x)= \partial_\mu^x  A_\nu(x)-\partial_\nu^x  A_\mu(x).
\end{eqnarray}
It should be noted that the derivative $\partial_x$ with respect to $x$ in triangle operator $\Delta$ only acts on the electromagnetic field tensor.

We can regard the Wigner function $W(x,p)$  as  one-body function while  $C^\mu(x,p)$ and $\Delta C^\mu(x,p)$ as two-body functions. Thus
the Wigner equation (\ref{W-eq-full}) shows that the equation of  one-body function depends on the two-body function, which is so-called the BBGKY  hierarchy\cite{Bogolyubov:1946}.
Within the mean field approximation, we can pull the tensor $ F_{\nu\mu}\left(x\right)$ out of the ensemble average and  the BBGKY  hierarchy truncates at the one-body function $W(x,p)$
with the result of  the quantum Vlasov equation.
For the general quantum fields, the function $C^\mu(x,p)$ and $\Delta C^\mu(x,p)$ will lead to  collision  terms, which are the main subject of our present work as indicated from the title of this paper.  However, it should be  emphasized that $C^\mu(x,p)$ and $\Delta C^\mu(x,p)$ include not only collisions terms but also all possible terms, such as mean field terms and so on. Hence, in our present work, the collision terms denote the full terms $C^\mu(x,p)$ and $\Delta C^\mu(x,p)$.   We will demonstrate that this general collision terms
must satisfy some self-consistent constraints determined by the Wigner equation itself. From now on, we will simply name  $C^\mu(x,p)$
and $\Delta C^\mu(x,p)$ or the associated function as the collision function.

The Wigner function  can be expanded in the 16 covariant matrices $1$, $i\gamma^5$, $\gamma_\nu$, $\gamma^5\gamma_\nu$, and
 $\sigma_{\mu\nu}=i[\gamma_\mu,\gamma_\nu]/2$, i.e.,
\begin{eqnarray}
\label{decomposition-W}
 W&=&\frac{1}{4}\left[\mathscr{F}+i\gamma^5\mathscr{P}+\gamma_\nu \mathscr{V}^\nu +\gamma^5\gamma_\nu \mathscr{A}^\nu
+\frac{1}{2}\sigma_{\mu\nu}\mathscr{S}^{\mu\nu}\right],
\end{eqnarray}
with the real coefficients representing  scalar $\mathscr{F}$, pseudoscalar $\mathscr{P}$, vector $\mathscr{V}^{\nu}$, axial vector $\mathscr{A}^{\nu}$ and antisymmetric tensor $\mathscr{S}^{\mu\nu}$ components, respectively.
In general, the   quantum kinetic theory for the fermions with arbitrary  mass might exhibit very different forms if we choose different
Wigner functions as independent distribution functions\cite{Gao:2019znl,Weickgenannt:2019dks,Hattori:2019ahi,Wang:2019moi,Ma:2022ins}.
In this work, we will follow the formalism of generalized chiral kinetic theory (GCKT)  derived in Ref.\cite{Ma:2022ins}. The GCKT can be reduced into the chiral kinetic theory with a smooth transition from massive to massless  fermions.
In GCKT, we  need introduce chiral Wigner function via
\begin{eqnarray}
\label{Js}
\mathscr{J}^\nu_s = \frac{1}{2}\left(\mathscr{V}^\nu + s \mathscr{A}^\nu\right),
\end{eqnarray}
where $s = +1/ -1$ denotes the chirality with right-handed/left-handed component.
Similarly, we can expand the collision functions as
\begin{eqnarray}
\label{decomposition-C}
{C}^\mu&=&\frac{1}{4}\left[\mathscr{C}^\mu+i\gamma^5\mathscr{C}^\mu_{5}+\gamma_\nu \mathscr{C}^{\mu\nu} +\gamma^5\gamma_\nu \mathscr{C}^{\mu\nu}_{5}
+\frac{1}{2}\sigma_{\alpha\beta}\mathscr{C}^{\mu\alpha\beta}\right],\\
\Delta {C}^\mu &=&\frac{1}{4}\left[\Delta\mathscr{C}^\mu +i\gamma^5\Delta\mathscr{C}^\mu_{5}+\gamma_\nu \Delta\mathscr{C}^{\mu\nu}
 +\gamma^5\gamma_\nu \Delta\mathscr{C}^{\mu\nu}_{5}
+\frac{1}{2}\sigma_{\alpha\beta}\Delta\mathscr{C}^{\mu\alpha\beta}\right],
\end{eqnarray}
with corresponding chiral collision functions

\begin{eqnarray}
\label{Cs}
\ \  \mathscr{C}^{\mu\nu}_s = \frac{1}{2}\left(\mathscr{C}^{\mu\nu} + s \mathscr{C}^{\mu\nu}_5\right),\ \ \Delta\mathscr{C}^{\mu\nu}_s = \frac{1}{2}\left(\Delta\mathscr{C}^{\mu\nu} + s \Delta\mathscr{C}^{\mu\nu}_5\right).
\end{eqnarray}
With these decomposition and chiral functions, the Wigner equation can be cast into

\begin{eqnarray}
\label{Js-cs-s}
m \mathscr{F} &=& 2p _\mu\mathscr{J}^\mu_s +  \hbar\Delta\mathscr{C}^\mu_{ s, \mu} ,
\\
\label{Js-eq}
- s m \mathscr{P} &=&\hbar\left( \partial_\mu^x \mathscr{J}^\mu_s  +  \mathscr{C}^\mu_{ s, \mu}\right),
\\
\label{Js-cs-t}
\frac{1}{2}m\epsilon_{\mu\nu\alpha\beta}\mathscr{S}^{\alpha\beta}
&=&\hbar \epsilon_{\mu\nu\alpha\beta}\left(\partial^\alpha_x \mathscr{J}^\beta_s +  \mathscr{C}^{\alpha\beta}_{ s}\right)\nonumber\\
& &+ s\left[2( p_\mu \mathscr{J}_{s,\nu}-p_\nu \mathscr{J}_{s,\mu}) + \hbar \left(\Delta\mathscr{C}_{s,\mu\nu}- \Delta\mathscr{C}_{s,\nu\mu}\right)\right] ,\\
\label{S-cs-s}
2m \sum_s\mathscr{J}_{s,\mu}  &=& 2 p_\mu \mathscr{F} + \hbar\Delta \mathscr{C}_{\mu}
+  \hbar\left( \partial^\nu_x \mathscr{S}_{\mu\nu} + \mathscr{C}^\nu_{\ \mu\nu}\right),\\
\label{P-eq}
-2m \sum_s s \mathscr{J}_{s,\mu} &=&\epsilon_{\mu\nu\alpha\beta}
\left(p^\nu\mathscr{S}^{\alpha\beta} +\frac{\hbar}{2}\Delta \mathscr{C}^{\nu\alpha\beta}\right)
- \hbar \left( \partial_\mu^x \mathscr{P} + \mathscr{C}_{ 5,\mu} \right),\\
\label{F-eq}
0  &=&\hbar \left(\partial_\mu^x \mathscr{F}+ \mathscr{C}_{\mu} \right)
-  2p^\nu \mathscr{S}_{\mu\nu} -\hbar \Delta\mathscr{C}^\nu_{\ \mu\nu} ,\\
\label{S-cs-t}
0 &=&\frac{1}{2}\hbar \epsilon_{\mu\nu\alpha\beta}
\left(\partial^\nu_x \mathscr{S}^{\alpha\beta} + \mathscr{C}^{\nu\alpha\beta} \right)
+ 2  p_\mu \mathscr{P}+ \hbar\Delta\mathscr{C}_{5,\mu} ,
\end{eqnarray}
where  we have recovered the $\hbar$ dependence so that we can make semiclassical expansion in the next section.
\section{Self-consistent constraints on  collision terms }
The Wigner equations given in Eqs.(\ref{Js-cs-s})-(\ref{S-cs-t}) are very complicated and totally coupled with each other.
It has been verified  in \cite{Gao:2018wmr,Gao:2019znl,Ma:2022ins} that these equations under mean field approximation  can be reduced to very simple form, in which only very few Wigner functions and equations are independent.  Some other Wigner functions can be derivative directly from the independent functions we chosen  and some Wigner  equations are fulfilled automatically and  redundant.  In this section, we will generalize this disentangling method   in mean field approximation to  general quantum field here.
In order to achieve this goal, we resort to the semiclassical $\hbar$ expansion. Besides  the explicit $\hbar$ dependence shown in Eqs.(\ref{Js-cs-s})-(\ref{S-cs-t}), the collision functions
also have explicit $\hbar$ expansion when we recover the $\hbar$ dependence after we  replace $\Delta$  by $\hbar \Delta$.
In current work, we will restrict ourselves to the  Wigner equation up to the first order. Then we only need to keep the leading contribution from $\Delta$ expansion:
\begin{eqnarray}
\label{C-up-2}
{C}^\mu_{ab} &=&\partial_\nu^p\int \frac{d^4y}{(2\pi)^4}e^{-ip\cdot y}
\langle  \bar\psi_b(x)e^{\frac{y}{2}\cdot D^\dagger} F^{\nu\mu}\left(x\right) e^{-\frac{y}{2}\cdot D}\psi_a(x)\rangle,\\
\Delta{C}^\mu_{ab} &=&\frac{\hbar}{6}\partial_\lambda^p \partial_\nu^p\int \frac{d^4y}{(2\pi)^4}e^{-ip\cdot y}
\langle  \bar\psi_b(x)e^{\frac{y}{2}\cdot D^\dagger} \partial_\lambda F^{\nu\mu}\left(x\right) e^{-\frac{y}{2}\cdot D}\psi_a(x)\rangle.
\end{eqnarray}
According to the decomposition in Eq.(\ref{decomposition-C}),  we have the  expressions up to this order,
\begin{eqnarray}
\label{def1}
 {\mathscr{C}^{\nu}} &\equiv&
\partial_\lambda^p \int\frac{d^4 y}{(2\pi)^4} e^{-ip\cdot y}
\left\langle \bar\psi e^{\frac{y}{2}\cdot D^\dagger}  F^{\nu\lambda}
e^{-\frac{y}{2}\cdot D}\psi\right\rangle,\\
\label{def2}
 {\mathscr{C}^{\nu}_5} &\equiv&
-i\partial_\lambda^p \int\frac{d^4 y}{(2\pi)^4} e^{-ip\cdot y}
\left\langle \bar\psi e^{\frac{y}{2}\cdot D^\dagger}  F^{\nu\lambda}\gamma^5
e^{-\frac{y}{2}\cdot D}\psi\right\rangle,\\
\label{def3}
\mathscr{C}^{\mu\nu}_s &\equiv&
\frac{1}{2}\partial_\lambda^p \int\frac{d^4 y}{(2\pi)^4} e^{-ip\cdot y}
\left\langle \bar\psi e^{\frac{y}{2}\cdot D^\dagger}  F^{\mu\lambda}\gamma^\nu(1 + s\gamma^5)
e^{-\frac{y}{2}\cdot D}\psi \right\rangle,\\
\label{def4}
 {\mathscr{C}^{\nu\alpha\beta}} &\equiv&
\partial_\lambda^p \int\frac{d^4 y}{(2\pi)^4} e^{-ip\cdot y}
\left\langle \bar\psi e^{\frac{y}{2}\cdot D^\dagger}  F^{\nu\lambda} \sigma^{\alpha\beta}
e^{-\frac{y}{2}\cdot D}\psi \right\rangle,
\end{eqnarray}
\begin{eqnarray}
\label{def5}
\Delta {\mathscr{C}^{\nu}} &\equiv&
\frac{\hbar}{6}\partial_\lambda^p \partial_\kappa^p\int\frac{d^4 y}{(2\pi)^4} e^{-ip\cdot y}
\left\langle \bar\psi e^{\frac{y}{2}\cdot D^\dagger} \partial^\kappa_x  F^{\nu\lambda}
e^{-\frac{y}{2}\cdot D}\psi \right\rangle,\\
\label{def6}
\Delta {\mathscr{C}^{\nu}_5} &\equiv&
-\frac{i\hbar}{6}\partial_\lambda^p \partial_\kappa^p \int\frac{d^4 y}{(2\pi)^4} e^{-ip\cdot y}
\left\langle \bar\psi e^{\frac{y}{2}\cdot D^\dagger} \partial^\kappa_x  F^{\nu\lambda}\gamma^5
e^{-\frac{y}{2}\cdot D}\psi \right\rangle,\\
\label{def7}
\Delta\mathscr{C}^{\mu\nu}_s &\equiv&
\frac{\hbar}{12}\partial_\lambda^p \partial_\kappa^p\int\frac{d^4 y}{(2\pi)^4} e^{-ip\cdot y}
\left\langle \bar\psi e^{\frac{y}{2}\cdot D^\dagger} \partial^\kappa_x  F^{\mu\lambda}\gamma^\nu(1 + s\gamma^5)
e^{-\frac{y}{2}\cdot D}\psi \right\rangle,\\
\label{def8}
\Delta {\mathscr{C}^{\nu\alpha\beta}} &\equiv&
\frac{\hbar}{6}\partial_\lambda^p\partial_\kappa^p \int\frac{d^4 y}{(2\pi)^4} e^{-ip\cdot y}
\left\langle \bar\psi e^{\frac{y}{2}\cdot D^\dagger} \partial^\kappa_x  F^{\nu\lambda} \sigma^{\alpha\beta}
e^{-\frac{y}{2}\cdot D}\psi \right\rangle.
\end{eqnarray}
where we have suppressed the arguments $x$ of the fields $\bar \psi$, $\psi$, and $ F^{\nu\lambda}$ for simplicity of notations.
We should note that even only leading term defined above can also contribute at any higher order which is implicit in the ensemble average of the operators, i.e.,
\begin{eqnarray}
\label{C-up-2}
\mathscr{C}^\mu_{ab} = \sum_{k=0}^\infty \hbar^k \mathscr{C}^{(k)\mu}_{ab},\ \ \
\Delta\mathscr{C}^\mu_{ab} = \sum_{k=0}^\infty \hbar^{k+1}\Delta \mathscr{C}^{(k)\mu}_{ab},
\end{eqnarray}
just as we do for the Wigner function
\begin{eqnarray}
 W (x,p) =\sum_{k=0}^\infty \hbar^k  W^{(k)}(x,p).
\end{eqnarray}

In order to disentangle the Wigner equations further, we introduce time-like 4-vector $n_\mu$ with normalization condition $n^2=1$.
In this work, we will assume that $n_\mu$ can be a function of coordinates $x^\mu$.
Then we can  decompose any 4-vector $X^\mu$ as
\begin{equation}
X^\mu=X_n n^\mu + \bar X^\mu,
\end{equation}
where $X_n=X\cdot n$ and $\bar X^\mu = \Delta^{\mu\nu}X_\nu$ with $\Delta^{\mu\nu}=g^{\mu\nu}-n^\mu n^\nu$.
We can also decompose the antisymmetric tensor as
\begin{eqnarray}
\label{S-KM}
\mathscr{S}^{\mu\nu}&=& \mathscr{K}^\mu n^\nu - \mathscr{K}^\nu n^\mu +\epsilon^{\mu\nu\rho\sigma}n_\rho \mathscr{M}_{\sigma},
\end{eqnarray}
with the evident relations
\begin{eqnarray}
\label{K-M}
\mathscr{K}^\mu &=& \mathscr{S}^{\mu\nu} n_\nu ,\ \ \mathscr{M}^\mu=\frac{1}{2}\epsilon^{\mu\nu\rho\sigma}n_\nu \mathscr{S}_{\rho\sigma}.
\end{eqnarray}
It is also convenient to  define the totally space-like antisymmetric tensor as
\begin{eqnarray}
\bar\epsilon_{\mu\alpha\beta}=\epsilon_{\mu\nu\alpha\beta}n^\nu.
\end{eqnarray}
\subsection{Zeroth-order result}
At zeroth order, all the collision functions vanish in the Wigner equations  Eqs.(\ref{Js-cs-s})-(\ref{S-cs-t}).  After the time-like and space-like decomposition according to the time-like vector $n^\mu$ for the Wigner functions and equations, the zeroth-order result can be presented as the following
\begin{eqnarray}
\label{Js-cs-s-0-nbar}
m {\mathscr{F}}^{(0)} &=& 2 \left(p_n {\mathscr{J}}^{(0)}_{s,n} +  \bar p_\mu \bar{\mathscr{J}}^{(0)\mu}_s\right),
\\
\label{Js-eq-0-nbar}
- s m {\mathscr{P}}^{(0)} &=&0,
\\
\label{Js-cs-t-0-n}
m{\mathscr{M}}^{(0)}_\mu
&=& 2 s \left(\bar p_\mu {\mathscr{J}}_{s,n}^{(0)} -  p_n \bar{\mathscr{J}}_{s,\mu}^{(0)}\right),\\
\label{Js-cs-t-0-bar}
m\bar\epsilon_{\mu\nu\alpha} {\mathscr{K}}^{(0)\alpha}
&=& 2 s \left(\bar p_\mu \bar{\mathscr{J}}_{s,\nu}^{(0)} -  \bar p_\nu \bar{\mathscr{J}}_{s,\mu}^{(0)}\right),\\
\label{S-cs-s-0-n}
2m \sum_s{\mathscr{J}}_{s,n}^{(0)}  &=& 2 p_n {\mathscr{F}}^{(0)},
\\
\label{S-cs-s-0-bar}
2m \sum_s\bar{\mathscr{J}}_{s,\mu}^{(0)}  &=& 2\bar p_\mu {\mathscr{F}}^{(0)},
\\
\label{P-eq-0-n}
-2m \sum_s s {\mathscr{J}}_{s,n}^{(0)} &=&-2\bar p^\mu {\mathscr{M}}^{(0)}_\mu ,
\\
\label{P-eq-0-bar}
-2m \sum_s s \bar{\mathscr{J}}_{s,\mu}^{(0)} &=&
2\left(p_n \mathscr{M}^{(0)}_\mu +\bar\epsilon_{\mu\nu\alpha}\bar p^\nu \mathscr{K}^{(0)\alpha}\right),
\\
\label{F-eq-0-n}
0  &=& -  2p^\nu{\mathscr{K}}_{\nu}^{(0)},
\\
\label{F-eq-0-bar}
0  &=& -  2\left(p_n \mathscr{K}^{(0)}_\mu -\bar\epsilon_{\mu\nu\alpha}\bar p^\nu \mathscr{M}^{(0)\alpha}\right),
\\
\label{S-cs-t-0-n}
0 &=&2 p_n {\mathscr{P}}^{(0)},
\\
\label{S-cs-t-0-bar}
0 &=&2 \bar p_\mu {\mathscr{P}}^{(0)}.
\end{eqnarray}
The sequence of the equations  listed above is in correspondence with the one  in  Eqs.(\ref{Js-cs-s})-(\ref{S-cs-t}).
The equation (\ref{Js-cs-t-0-n}) is from the tensor equation (\ref{Js-cs-t}) with the time-like component $\nu$ and space-like component $\mu$
while the equation (\ref{Js-cs-t-0-bar}) is from the space-like and space-like components for both $\mu$ and $\nu$.
The equations (\ref{S-cs-s-0-n})/(\ref{S-cs-s-0-bar}), (\ref{P-eq-0-n})/(\ref{P-eq-0-bar}),  (\ref{F-eq-0-n})/(\ref{F-eq-0-bar}), and  (\ref{S-cs-t-0-n})/(\ref{S-cs-t-0-bar})
  are from the time-like/space-like components of Eqs.(\ref{S-cs-s}),(\ref{P-eq}), (\ref{F-eq}), and (\ref{S-cs-t}), respectively

We will choose ${\mathscr{J}}_{s,n}$ and $ \mathscr{M}_\mu $ as the independent Wigner functions and the other Wigner functions as the derived functions. From the Eqs.(\ref{Js-cs-t-0-n}), (\ref{S-cs-s-0-n}),(\ref{F-eq-0-bar}) and (\ref{S-cs-t-0-n}), we can express the Wigner functions $\bar{\mathscr{J}}_{s,\mu}^{(0)}$,
${\mathscr{F}}^{(0)}$, $\mathscr{K}^{(0)}_\mu $ and $ {\mathscr{P}}^{(0)}$ in terms of independent functions ${\mathscr{J}}_{s,n}^{(0)}$ and ${\mathscr{M}}^{(0)}_\mu$, respectively,
\begin{eqnarray}
\label{Jbar-0}
\bar{\mathscr{J}}_{s,\mu}^{(0)}
&=& \frac{\bar p_\mu}{p_n} {\mathscr{J}}_{s,n}^{(0)}-\frac{s m}{2p_n}{\mathscr{M}}^{(0)}_\mu,\\
\label{F-0}
{\mathscr{F}}^{(0)}   &=&  \frac{ m}{p_n} \sum_s{\mathscr{J}}_{s,n}^{(0)} ,
\\
\label{K-0}
\mathscr{K}^{(0)\mu}  &=& \frac{1}{p_n}\bar\epsilon^{\mu\nu\alpha}\bar p_\nu \mathscr{M}_\alpha^{(0)},
\\
\label{P-0}
{\mathscr{P}}^{(0)} &=& 0.
\end{eqnarray}
Substituting Eq.(\ref{K-0}) into Eq.(\ref{S-KM}), we can obtain the antisymmetric Wigner function
\begin{eqnarray}
\label{S-M-0}
\mathscr{S}^{(0)\mu\nu}&=&\frac{1}{p_n} \epsilon^{\mu\nu\alpha\beta}p_\alpha \mathscr{M}_{\beta}^{(0)}.
\end{eqnarray}
The  Eq.(\ref{P-eq-0-n}) gives the constraint condition for  ${\mathscr{M}}^{(0)}_\mu$
\begin{eqnarray}
\label{M-cs-0}
\bar p^\mu {\mathscr{M}}^{(0)}_\mu  &=& m \sum_s s {\mathscr{J}}_{s,n}^{(0)}.
\end{eqnarray}
This constraint and the one $n^\mu{\mathscr{M}}^{(0)}_\mu=0 $ directly from the definition (\ref{K-M}) means only two components are independent for  ${\mathscr{M}}^{(0)}_\mu$.
It is convenient to decompose $\mathscr{M}^{(0)}_\mu$ into the longitudinal and transverse parts with respect to the momentum $\bar p^\mu$
\begin{eqnarray}
\mathscr{M}^{(0)}_{\mu}= \mathscr{M}^{(0)}_{\parallel\mu}+ \mathscr{M}^{(0)}_{\perp\mu}, \ \ \ \textrm{with}\ \ \
\mathscr{M}^{(0)}_{\parallel \mu}=  \frac{m \bar p_\mu}{\bar p^2}  \sum_s s {\mathscr{J}}_{s,n}^{(0)}  \ \ \textrm{and}\ \ \bar p^\mu \mathscr{M}^{(0)}_{\perp\mu}=0.
\end{eqnarray}
We note that the longitudinal part   $\mathscr{M}^{\mu}_\parallel$  along the direction of $\bar p^\mu$ is totally determined by the function
$ {\mathscr{J}}_{s,n}^{(0)}$  and  the independent part
is only transverse component $\mathscr{M}^{\mu}_\perp$.

Substituting the expressions given in (\ref{Jbar-0})-(\ref{P-0}) into Eqs.(\ref{Js-cs-s-0-nbar}) and (\ref{P-eq-0-bar}) and using the constraint (\ref{M-cs-0}) , we obtain the on-shell conditions for ${\mathscr{J}}_{s,n}^{(0)}$ and ${\mathscr{M}}^{(0)}_\mu$,
respectively,
\begin{eqnarray}
\label{Jn-os-0}
\left(p^2-m^2\right)\frac{ {\mathscr{J}}_{s,n}^{(0)}}{p_n} &=& 0,\\
\label{M-os-0}
\left(p^2-m^2\right)\frac{ {\mathscr{M}}_{\mu}^{(0)}}{p_n} &=& 0.
\end{eqnarray}
The expressions take the general form of
\begin{eqnarray}
\label{Jn-os-0-r}
 {\mathscr{J}}_{s,n}^{(0)} &=& p_n {\mathcal{J}}_{s,n}^{(0)}\delta(p^2-m^2) ,\\
\label{M-os-0-r}
 {\mathscr{M}}_{\mu}^{(0)} &=& p_n {\mathcal{M}}_{\mu}^{(0)}\delta(p^2-m^2),
\end{eqnarray}
where ${\mathcal{J}}_{s,n}^{(0)}$ and ${\mathcal{M}}_{\mu}^{(0)}$ are both regular functions of $x^\mu$ and $p^\mu$  at $p^2-m^2=0$.

Once the equations (\ref{Jbar-0})-(\ref{P-0}) hold , it is trivial to verify that the Eqs.(\ref{Js-eq-0-nbar}), (\ref{Js-cs-t-0-bar}), (\ref{S-cs-s-0-bar}), (\ref{F-eq-0-n}) and (\ref{S-cs-t-0-bar}) are all satisfied automatically.

At zeroth order, we have no  either constraints for the collision function or the kinetic equation for independent Wigner functions ${\mathscr{J}}_{s,n}^{(0)}$ and ${\mathscr{M}}^{(0)}_\mu$.

\subsection{First-order result}
The collision terms and kinetic equation begin to appear at first order. Similar to the procedure we carried out  at  zeroth order,  the first-order Wigner equations can be obtained from  Eqs.(\ref{Js-cs-s})-(\ref{S-cs-t})
\begin{eqnarray}
\label{Js-cs-s-1-nbar}
m {\mathscr{F}}^{(1)} &=& 2 \left(p_n {\mathscr{J}}^{(1)}_{s,n} +  \bar p_\mu \bar{\mathscr{J}}^{(1)\mu}_s\right),
\\
\label{Js-eq-1-nbar}
- s m {\mathscr{P}}^{(1)} &=&\partial^x_\mu {\mathscr{J}}^{(0)\mu}_s
+\mathscr{C}^{(0)\mu}_{s,\mu},
\\
\label{Js-cs-t-1-n}
m{\mathscr{M}}^{(1)}_\mu
&=& 2 s \left(\bar p_\mu {\mathscr{J}}_{s,n}^{(1)} -  p_n \bar{\mathscr{J}}_{s,\mu}^{(1)}\right)
+\bar\epsilon_{\mu\alpha\beta}\partial_x^{\alpha}{\mathscr{J}}^{(0)\beta}_s
+\bar\epsilon_{\mu\alpha\beta} \mathscr{C}^{(0)\alpha\beta}_{s},
\\
\label{Js-cs-t-1-bar}
m\bar\epsilon_{\mu\nu\alpha} {\mathscr{K}}^{(1)\alpha}
&=& 2 s \left(\bar p_\mu \bar{\mathscr{J}}_{s,\nu}^{(1)} -  \bar p_\nu \bar{\mathscr{J}}_{s,\mu}^{(1)}\right)
+\bar\epsilon_{\mu\nu\alpha}n_\beta\left(\partial_x^{\alpha}{\mathscr{J}}^{(0)\beta}_s
- \partial_x^{\beta}{\mathscr{J}}^{(0)\alpha}_s\right)\nonumber\\
& &+\bar\epsilon_{\mu\nu\alpha}n_\beta \left(\mathscr{C}^{(0)\alpha\beta}_{s}
-\mathscr{C}^{(0)\beta\alpha}_{s}\right),
\end{eqnarray}
\begin{eqnarray}
\label{S-cs-s-1-n}
2m \sum_s{\mathscr{J}}_{s,n}^{(1)}  &=& 2 p_n {\mathscr{F}}^{(1)}
+ n^\mu\partial_x^{\nu}{\mathscr{S}}_{\mu\nu}^{(0)}
+n^\mu{\mathscr{C}^{(0)\nu}}_{\mu\nu},
\\
\label{S-cs-s-1-bar}
2m \sum_s\bar{\mathscr{J}}_{s,\mu}^{(1)}  &=& 2\bar p_\mu {\mathscr{F}}^{(1)}
+\Delta_{\mu}^\lambda \partial_x^{\nu}{\mathscr{S}}_{\lambda\nu}^{(0)}
+\Delta_{\mu}^\lambda{\mathscr{C}^{(0)\nu}}_{\lambda\nu},
\\
\label{P-eq-1-n}
-2m \sum_s s {\mathscr{J}}_{s,n}^{(1)} &=&-2\bar p_\mu {\mathscr{M}}^{(1)\mu }
-n^\mu \partial^x_\mu {\mathscr{P}}^{(0)} -\mathscr{C}^{(0)}_{5, n},
\\
\label{P-eq-1-bar}
-2m \sum_s s \bar{\mathscr{J}}_{s,\mu}^{(1)} &=&
2\left(p_n \mathscr{M}^{(1)}_\mu +\bar\epsilon_{\mu\nu\alpha}\bar p^\nu \mathscr{K}^{(1)\alpha}\right)
-\Delta_\mu^\lambda \partial^x_\lambda {\mathscr{P}}^{(0)} -\bar{ \mathscr{C}^{(0)}}_{5,\mu},
\end{eqnarray}
\begin{eqnarray}
\label{F-eq-1-n}
0  &=&   2p^\nu{\mathscr{K}}_{\nu}^{(1)}
+n^\mu\partial_\mu^x {\mathscr{F}}^{(0)} + \mathscr{C}^{(0)}_{n},
\\
\label{F-eq-1-bar}
0  &=& -  2\left(p_n \mathscr{K}^{(1)}_\mu -\bar\epsilon_{\mu\nu\alpha}\bar p^\nu \mathscr{M}^{(1)\alpha}\right)
+\Delta_\mu^\lambda\partial_\lambda^x {\mathscr{F}}^{(0)}
+ \bar{\mathscr{C}}^{(0)}_{\mu},
\\
\label{S-cs-t-1-n}
0 &=&2 p_n {\mathscr{P}}^{(1)}
-\frac{1}{2}\bar \epsilon_{\nu\alpha\beta}\partial^{\nu}_x {\mathscr{S}}^{(0)\alpha\beta}
-\frac{1}{2}\bar \epsilon_{\nu\alpha\beta}\mathscr{C}^{(0)\nu\alpha\beta},
\\
\label{S-cs-t-1-bar}
0 &=&2 \bar p_\mu {\mathscr{P}}^{(1)}
+\frac{1}{2}\Delta_\mu^\lambda \epsilon_{\lambda\nu\alpha\beta}\partial^{\nu}_x {\mathscr{S}}^{(0)\alpha\beta}
+\frac{1}{2} \Delta_\mu^\lambda  \epsilon_{\lambda\nu\alpha\beta}\mathscr{C}^{(0)\nu\alpha\beta}.
\end{eqnarray}
From the Eqs.(\ref{Js-cs-t-1-n}), (\ref{S-cs-s-1-n}),(\ref{F-eq-1-bar}), and (\ref{S-cs-t-1-n}),  we can express the functions
  $\bar{\mathscr{J}}_{s,\mu}^{(1)}$, ${\mathscr{F}}^{(1)}$, $\mathscr{K}^{(1)}_\mu $ and $ {\mathscr{P}}^{(1)}$ in terms of independent functions ${\mathscr{J}}_{s,n}^{(1)}$ and ${\mathscr{M}}^{(1)}_\mu$, respectively,
\begin{eqnarray}
\label{Jbar-1}
\bar{\mathscr{J}}_{s,\mu}^{(1)}
&=& \frac{\bar p_\mu}{p_n} {\mathscr{J}}_{s,n}^{(1)}-\frac{s m}{2p_n}{\mathscr{M}}^{(1)}_\mu
+\frac{s}{2p_n}\bar\epsilon_{\mu\alpha\beta}\partial_x^{\alpha}{\mathscr{J}}^{(0)\beta}_s
+\frac{s}{2p_n}\bar\epsilon_{\mu\alpha\beta} \mathscr{C}^{(0)\alpha\beta}_{s},\\
\label{F-1}
{\mathscr{F}}^{(1)}   &=&  \frac{ m}{p_n} \sum_s{\mathscr{J}}_{s,n}^{(1)}
-\frac{ 1}{2p_n}  n^\mu\partial_x^{\nu}{\mathscr{S}}_{\mu\nu}^{(0)}
-\frac{ 1}{2p_n} n^\mu{\mathscr{C}^{(0)\nu}}_{\mu\nu},
\\
\label{K-1}
\mathscr{K}^{(1)\mu}  &=& \frac{1}{p_n}\bar\epsilon^{\mu\nu\alpha}\bar p_\nu \mathscr{M}_\alpha^{(1)}
+\frac{ 1}{2p_n}\Delta^{\mu\lambda}\partial_\lambda^x {\mathscr{F}}^{(0)}
+\frac{ 1}{2p_n} \bar{\mathscr{C}}^{(0)\mu},\\
\label{P-1}
{\mathscr{P}}^{(1)} &=& \frac{1}{4p_n}\bar \epsilon_{\nu\alpha\beta}\partial^{\nu}_x {\mathscr{S}}^{(0)\alpha\beta}
+\frac{1}{4p_n}\bar \epsilon_{\nu\alpha\beta}\mathscr{C}^{(0)\nu\alpha\beta}.
\end{eqnarray}
Substituting Eq.(\ref{K-1}) into Eq.(\ref{S-KM}), we can obtain the antisymmetric Wigner function
\begin{eqnarray}
\label{S-M-1}
\mathscr{S}^{(1)\mu\nu}=\frac{1}{p_n} \epsilon^{\mu\nu\alpha\beta}p_\alpha \mathscr{M}_{\beta}^{(1)}
+\frac{1}{2p_n}\left(\Delta^{\mu\lambda} n^\nu - \Delta^{\nu\lambda} n^\mu\right)\left(\partial_\lambda^x {\mathscr{F}}^{(0)}
+ \mathscr{C}^{(0)}_{\lambda}\right).
\end{eqnarray}
The difference from the zeroth-order results is that the collision  functions have been involved at first order.
The  Eq.(\ref{P-eq-1-n}) gives the longitudinal constraint condition for ${\mathscr{M}}^{(1)}_\mu$
\begin{eqnarray}
\label{M-cs-1}
\bar p_\mu {\mathscr{M}}^{(1)\mu}  &=& m \sum_s s {\mathscr{J}}_{s,n}^{(1)} - \frac{1}{2}\mathscr{C}^{(0)}_{5, n}.
\end{eqnarray}
Just like at zeroth order, we can  decompose $\mathscr{M}^{(1)}_\mu$ into the longitudinal and transverse parts
$\mathscr{M}^{(1)}_{\mu}= \mathscr{M}^{(1)}_{\parallel\mu}+ \mathscr{M}^{(1)}_{\perp\mu}$ with respect to momentum $\bar p^\mu$
and only transverse component $\mathscr{M}^{\mu}_\perp$ is independent.

The  Eqs.(\ref{Js-cs-s-1-nbar}) and (\ref{P-eq-1-bar}) together with Eq.(\ref{M-cs-1}) give the modification to the  on-shell conditions of ${\mathscr{J}}_{s,n}^{(1)}$ and ${\mathscr{M}}^{(1)}_\mu$, respectively,
\begin{eqnarray}
\label{Jn-os-1}
\left(p^2-m^2\right)\frac{ {\mathscr{J}}_{s,n}^{(1)}}{p_n}
&=& -\frac{sm}{4p_n}\mathscr{C}^{(0)}_{5,n}
-\frac{s}{2p_n}\bar\epsilon_{\mu\alpha\beta}\bar p^\mu \mathscr{C}^{(0)\alpha\beta}_{s}
-\frac{ m}{4p_n} n^\mu{\mathscr{C}^{(0)\nu}}_{\mu\nu},
\\
\label{M-os-1}
\left(p^2-m^2\right)\frac{ {\mathscr{M}}_{\mu}^{(1)}}{p_n}
&=& \frac{1}{2}\bar{ \mathscr{C}}^{(0)}_{5,\mu}-\frac{\bar p_\mu}{2p_n}{ \mathscr{C}}^{(0)}_{5,n}
-\frac{ 1}{2p_n} \bar\epsilon_{\mu\nu\alpha}\bar p^\nu \bar{\mathscr{C}}^{(0)\alpha}
-\frac{m}{2p_n}\bar\epsilon_{\mu\alpha\beta}\sum_s \mathscr{C}^{(0)\alpha\beta}_{s}.
\end{eqnarray}
At zeroth order, the remaining Wigner equations are all satisfied automatically. At first order, these equations will result in kinetic equations or constraints on the collision terms.
For example, the Eqs.(\ref{Js-eq-1-nbar}) and (\ref{S-cs-t-1-bar}) give the quantum kinetic equations for the zeroth order Wigner functions ${\mathscr{J}}_{s,n}^{(0)}$ and ${\mathscr{M}}^{(0)}_\mu$, respectively,
\begin{eqnarray}
\label{Jsn-eq-0}
p^\mu \partial^x_\mu\left( \frac{\mathscr{J}_{s,n}^{(0)}}{p_n}\right)
&=&-\frac{ m s}{2p_n}p^\nu \left(\partial^x_\nu  n_\mu \right)\frac{{\mathscr{M}}^{(0)\mu} }{p_n}
-\mathscr{C}^{(0)\mu}_{s,\mu}
-\frac{ms}{4p_n}\bar \epsilon_{\nu\alpha\beta}\mathscr{C}^{(0)\nu\alpha\beta},\\
\label{M-eq-0}
p^\nu \partial^x_\nu \left(\frac{\mathscr{M}^{(0)\mu}}{p_n}\right)
&=&-\frac{1}{p_n} p^\mu p^\nu \left( \partial^x_\nu n_\lambda \right)\frac{\mathscr{M}^{(0)\lambda}}{p_n}
-\frac{1}{2}\left(\epsilon^{\mu\nu\alpha\beta} + \frac{p^\mu}{p_n}\bar\epsilon^{\nu\alpha\beta}\right)
\mathscr{C}^{(0)}_{\nu\alpha\beta}.
\end{eqnarray}

Substituting Eqs.(\ref{Jbar-1}) and (\ref{K-1}) into Eq.(\ref{Js-cs-t-1-bar}) and doing  some vector algebra  together with the kinetic equation (\ref{Jsn-eq-0}), we obtain
the constraint equation
\begin{eqnarray}
\label{collision-4-a}
\Delta_\alpha^\lambda\left(\frac{ m}{2} {\mathscr{C}}^{(0)\alpha}
+\frac{sm}{4}\epsilon^{\alpha\beta\rho\sigma} \mathscr{C}^{(0)}_{\beta\rho\sigma}\right)
&=&
\Delta_\alpha^\lambda\left[ p_\beta  \left( \mathscr{C}^{(0)\alpha\beta}_{s}- \mathscr{C}^{(0)\beta\alpha}_{s}\right)
+  p^\alpha \mathscr{C}^{(0)\beta}_{s,\beta}\right].
\end{eqnarray}
From the definitions (\ref{def1}-\ref{def8}), we notice that all these collision functions cannot depend on the auxiliary time-like vector $n^\mu$.
Hence the requirement that the constraint (\ref{collision-4-a}) hold for any $n^\mu$ lead to the following constraint condition
\begin{eqnarray}
\label{collision-4-a}
\frac{ m}{2} {\mathscr{C}}^{(0)\alpha}
+\frac{sm}{4}\epsilon^{\alpha\beta\rho\sigma} \mathscr{C}^{(0)}_{\beta\rho\sigma}
&=&
 p_\beta  \left( \mathscr{C}^{(0)\alpha\beta}_{s}- \mathscr{C}^{(0)\beta\alpha}_{s}\right)
+  p^\alpha \mathscr{C}^{(0)\beta}_{s,\beta}.
\end{eqnarray}
Substituting Eqs.(\ref{Jbar-1}) and (\ref{F-1}) into Eq.(\ref{S-cs-s-1-bar}) and using the kinetic equation (\ref{M-eq-0}) and the result (\ref{S-M-0}) give rise to
the constraint equation
\begin{eqnarray}
\label{collision-3-a-n}
m n^\nu \epsilon_{\mu\nu\alpha\beta}  \sum_s s\mathscr{C}^{(0)\alpha\beta}_{s}
= n^\nu\left(-  p_\mu {\mathscr{C}^{(0)\lambda}}_{\nu \lambda}
+p_\nu  {\mathscr{C}^{(0)\lambda}}_{\mu\lambda}
-\frac{1}{2}\epsilon_{\mu\nu\alpha\beta} p^\alpha\epsilon^{\beta\lambda\rho\sigma}
{\mathscr{C}^{(0)}}_{\lambda\rho\sigma}\right).
\end{eqnarray}
The fact that this constraint holds for any $n^\mu$ lead to more general constraint condition
\begin{eqnarray}
\label{collision-3-a}
m\epsilon_{\mu\nu\alpha\beta}  \sum_s s\mathscr{C}^{(0)\alpha\beta}_{s}
= -  p_\mu {\mathscr{C}^{(0)\lambda}}_{\nu \lambda}
+p_\nu  {\mathscr{C}^{(0)\lambda}}_{\mu\lambda}
-\frac{1}{2}\epsilon_{\mu\nu\alpha\beta} p^\alpha\epsilon^{\beta\lambda\rho\sigma}
{\mathscr{C}^{(0)}}_{\lambda\rho\sigma}.
\end{eqnarray}
Acting the operator $p^\nu\partial_\nu^x$  on the  Eq. (\ref{M-cs-0})  and using  Eqs.(\ref{Jsn-eq-0}) and (\ref{M-eq-0})   lead to
\begin{eqnarray}
\label{collision-2-a-n}
2m n^\mu p_\mu\sum_s s\mathscr{C}^{(0)\nu}_{s,\nu}
&=&n^\mu \left[p_\mu p^\lambda\epsilon_{\lambda\nu\alpha\beta}-\left(p^2-m^2\right)\epsilon_{\mu\nu\alpha\beta}\right]\mathscr{C}^{(0)\nu\alpha\beta}.
\end{eqnarray}
For the same reason as Eqs.(\ref{collision-3-a-n}), this will lead to more general constraint
\begin{eqnarray}
\label{collision-2-a}
2m  p_\mu\sum_s s\mathscr{C}^{(0)\nu}_{s,\nu}
&=& \left[p_\mu p^\lambda\epsilon_{\lambda\nu\alpha\beta}-\left(p^2-m^2\right)\epsilon_{\mu\nu\alpha\beta}\right]\mathscr{C}^{(0)\nu\alpha\beta}.
\end{eqnarray}
Plugging Eq.(\ref{K-1}) into Eq.(\ref{F-eq-1-n}) together with  Eqs.(\ref{F-0}) and (\ref{Jsn-eq-0})  lead to
\begin{eqnarray}
\label{collision-1-a}
m\sum_s \mathscr{C}^{(0)\mu}_{s,\mu}
&=& p^\mu \mathscr{C}^{(0)}_{\mu}.
\end{eqnarray}
Multiplying  Eq.(\ref{M-cs-1}) by $(p^2-m^2)$ and using the Eqs.(\ref{Jn-os-1}) and (\ref{M-os-1}) gives
\begin{eqnarray}
\label{collision-0-a}
p^\mu \mathscr{C}_{5,\mu}^{(0)}&=&0.
\end{eqnarray}
Acting the operator $p^\nu\partial_\nu^x$ on (\ref{Jn-os-0}) and multiplying (\ref{Jsn-eq-0}) by $(p^2-m^2)$ gives rise to
\begin{eqnarray}
\label{onshell-C-Jn}
0&=&\left(p^2-m^2\right)\left(p_\mu\mathscr{C}^{(0)\nu}_{s,\nu}-\frac{sm}{4}\epsilon_{\mu\nu\alpha\beta}\mathscr{C}^{(0)\nu\alpha\beta}\right).
\end{eqnarray}
Making similar manipulation on  Eqs.(\ref{M-os-0}) and (\ref{M-eq-0}) gives rise to
\begin{eqnarray}
\label{onshell-C-M}
0&=&\left(p^2-m^2\right)\left(p_\lambda\epsilon_{\mu\nu\alpha\beta}-p_\mu\epsilon_{\lambda\nu\alpha\beta}\right)\mathscr{C}^{(0)\nu\alpha\beta}.
\end{eqnarray}
To obtain the final results in Eqs.(\ref{onshell-C-Jn}) and (\ref{onshell-C-M}), we have used the property of the arbitrariness of $n^\mu$.
However, it is easy to verify that the constraint equation  (\ref{onshell-C-M}) can be derived from Eq.(\ref{collision-2-a}) while (\ref{onshell-C-Jn}) can be derived
from Eqs. (\ref{collision-1-a}), (\ref{collision-2-a}) and (\ref{collision-3-a}). Hence the final independent  constraint equations for the collision functions are given by
\begin{eqnarray}
\label{collision-0-f}
p^\mu \mathscr{C}_{5,\mu}^{(0)}&=&0,\\
\label{collision-1-f}
m\sum_s \mathscr{C}^{(0)\mu}_{s,\mu}
&=& p^\mu \mathscr{C}^{(0)}_{\mu},
\\
\label{collision-2-f}
2mp_\mu\sum_s s\mathscr{C}^{(0)\nu}_{s,\nu}
&=&\left[p_\mu p^\lambda\epsilon_{\lambda\nu\alpha\beta}-\left(p^2-m^2\right)\epsilon_{\mu\nu\alpha\beta}\right]\mathscr{C}^{(0)\nu\alpha\beta},
\\
\label{collision-3-f}
m \epsilon_{\mu\nu\alpha\beta}  \sum_s s\mathscr{C}^{(0)\alpha\beta}_{s}
&=& p_\nu  {\mathscr{C}^{(0)\lambda}}_{\mu\lambda}
-  p_\mu {\mathscr{C}^{(0)\lambda}}_{\nu \lambda}
-\frac{1}{2}\epsilon_{\mu\nu\alpha\beta} p^\alpha\epsilon^{\beta\lambda\rho\sigma}
{\mathscr{C}^{(0)}}_{\lambda\rho\sigma},\hspace{1cm}
\\
\label{collision-4-f}
\frac{ m}{2} {\mathscr{C}}^{(0)\alpha}
+\frac{sm}{4}\epsilon^{\alpha\beta\rho\sigma} \mathscr{C}^{(0)}_{\beta\rho\sigma}
&=&
p_\beta  \left( \mathscr{C}^{(0)\alpha\beta}_{s}- \mathscr{C}^{(0)\beta\alpha}_{s}\right)
+  p^\alpha \mathscr{C}^{(0)\beta}_{s,\beta}.
\end{eqnarray}

At first order, we have obtained the  constraints for the zeroth-order collision functions and  the  kinetic equation for zeroth-order independent Wigner functions ${\mathscr{J}}_{s,n}^{(0)}$ and ${\mathscr{M}}^{(0)}_\mu$. Since these constraint equations are derived only  from the consistency of the Wigner equations, we call them
self-consistent constraint equations.
\subsection{Second-order result}
In order to obtain the kinetic equation for the first-order independent Wigner functions ${\mathscr{J}}_{s,n}^{(1)}$ and ${\mathscr{M}}^{(1)}_\mu$,
we need to consider the Wigner equations at  second order.  We can write second-order Wigner equations in the similar form as   first order,
\begin{eqnarray}
\label{Js-cs-s-2-nbar}
m {\mathscr{F}}^{(2)} &=& 2 \left(p_n {\mathscr{J}}^{(2)}_{s,n} +  \bar p_\mu \bar{\mathscr{J}}^{(2)\mu}_s\right)
+\Delta\mathscr{C}_{s,\mu}^{(0)\mu},
\\
\label{Js-eq-2-nbar}
- s m {\mathscr{P}}^{(2)} &=&\partial^x_\mu {\mathscr{J}}^{(1)\mu}_s
+\tilde{\mathscr{C}}^{(1)\mu}_{s,\mu},
\\
\label{Js-cs-t-2-n}
m{\mathscr{M}}^{(2)}_\mu
&=& 2 s \left(\bar p_\mu {\mathscr{J}}_{s,n}^{(2)} -  p_n \bar{\mathscr{J}}_{s,\mu}^{(2)}\right)
+\bar\epsilon_{\mu\alpha\beta}\partial_x^{\alpha}{\mathscr{J}}^{(1)\beta}_s
+\bar\epsilon_{\mu\alpha\beta}\tilde{ \mathscr{C}}^{(1)\alpha\beta}_{s},
\\
\label{Js-cs-t-2-bar}
m\bar\epsilon_{\mu\nu\alpha} {\mathscr{K}}^{(2)\alpha}
&=& 2 s \left(\bar p_\mu \bar{\mathscr{J}}_{s,\nu}^{(2)} -  \bar p_\nu \bar{\mathscr{J}}_{s,\mu}^{(2)}\right)
+\bar\epsilon_{\mu\nu\alpha}n_\beta\left(\partial_x^{\alpha}{\mathscr{J}}^{(1)\beta}_s
- \partial_x^{\beta}{\mathscr{J}}^{(1)\alpha}_s\right)\nonumber\\
& &+\bar\epsilon_{\mu\nu\alpha}n_\beta \left(\tilde{\mathscr{C}}^{(1)\alpha\beta}_{s}
-\tilde{\mathscr{C}}^{(1)\beta\alpha}_{s}\right),
\end{eqnarray}
\begin{eqnarray}
\label{S-cs-s-2-n}
2m \sum_s{\mathscr{J}}_{s,n}^{(2)}  &=& 2 p_n {\mathscr{F}}^{(2)}
+ n^\mu\partial_x^{\nu}{\mathscr{S}}_{\mu\nu}^{(1)}
+n^\mu{\tilde{\mathscr{C}}^{(1)\nu}}_{\ \ \ \ \ \mu\nu} ,
\\
\label{S-cs-s-2-bar}
2m \sum_s\bar{\mathscr{J}}_{s,\mu}^{(2)}  &=& 2\bar p_\mu {\mathscr{F}}^{(2)}
+\Delta_{\mu}^\lambda \partial_x^{\nu}{\mathscr{S}}_{\lambda\nu}^{(1)}
+\Delta_{\mu}^\lambda{\tilde{\mathscr{C}}^{(1)\nu}}_{\ \ \ \ \ \lambda\nu},
\\
\label{P-eq-2-n}
-2m \sum_s s {\mathscr{J}}_{s,n}^{(2)} &=&-2\bar p_\mu {\mathscr{M}}^{(2)\mu }
-n^\mu \partial^x_\mu {\mathscr{P}}^{(1)} -\tilde{\mathscr{C}}^{(1)}_{5, n} ,
\\
\label{P-eq-2-bar}
-2m \sum_s s \bar{\mathscr{J}}_{s,\mu}^{(2)} &=&
2\left(p_n \mathscr{M}^{(2)}_\mu +\bar\epsilon_{\mu\nu\alpha}\bar p^\nu \mathscr{K}^{(2)\alpha}\right)
-\Delta_\mu^\lambda \partial^x_\lambda {\mathscr{P}}^{(1)}
-\Delta^\lambda_\mu \tilde{\mathscr{C}}^{(1)}_{5,\lambda} ,
\end{eqnarray}
\begin{eqnarray}
\label{F-eq-2-n}
0  &=&   2p^\nu{\mathscr{K}}_{\nu}^{(2)}
+n^\mu\partial_\mu^x {\mathscr{F}}^{(1)} + \tilde{\mathscr{C}}^{(1)}_{n},
\\
\label{F-eq-2-bar}
0  &=& -  2\left(p_n \mathscr{K}^{(2)\mu} -\bar\epsilon^{\mu\nu\alpha}\bar p_\nu \mathscr{M}_\alpha^{(2)}\right)
+\Delta_\mu^\lambda\partial_\lambda^x {\mathscr{F}}^{(1)}
+ \Delta_\mu^\lambda \tilde{\mathscr{C}}^{(1)}_{\lambda},
\\
\label{S-cs-t-2-n}
0 &=&2 p_n {\mathscr{P}}^{(2)}
-\frac{1}{2}\bar \epsilon_{\nu\alpha\beta}\partial^{\nu}_x {\mathscr{S}}^{(1)\alpha\beta}
-\frac{1}{2}\bar \epsilon_{\nu\alpha\beta}\tilde{\mathscr{C}}^{(1)\nu\alpha\beta},
\\
\label{S-cs-t-2-bar}
0 &=&2 \bar p_\mu {\mathscr{P}}^{(2)}
+\frac{1}{2}\Delta_\mu^\lambda \epsilon_{\lambda\nu\alpha\beta}\partial^{\nu}_x {\mathscr{S}}^{(1)\alpha\beta}
+\frac{1}{2} \Delta_\mu^\lambda  \epsilon_{\lambda\nu\alpha\beta}\tilde{\mathscr{C}}^{(1)\nu\alpha\beta},
\end{eqnarray}
where we have defined
\begin{eqnarray}
\label{tC-1}
\tilde{ \mathscr{C}}^{(1)}_\lambda &\equiv &  \mathscr{C}^{(1)}_\lambda - \Delta {\mathscr{C}^{(0)\nu}}_{\lambda\nu},\\
\label{tC-2}
\tilde{ \mathscr{C}}^{(1)}_{5,\lambda} &\equiv &  \mathscr{C}^{(1)}_{5,\lambda} -\frac{1}{2}\epsilon_{\lambda\nu\alpha\beta} \Delta \mathscr{C}^{(0)\nu\alpha\beta},\\
\label{tC-3}
\tilde{ \mathscr{C}}^{(1)\alpha\beta}_{s} &\equiv &
\mathscr{C}^{(1)\alpha\beta}_{s}
- \frac{s}{2} \epsilon^{\alpha\beta\rho\sigma}\Delta \mathscr{C}^{(0)}_{s,\rho\sigma},\\
\label{tC-4}
\tilde{ \mathscr{C}}^{(1)\nu\alpha\beta} &\equiv &  \mathscr{C}^{(1)\nu\alpha\beta}
+\frac{1}{3}\left( g^{\nu\beta}\Delta\mathscr{C}^{(0)\alpha} - g^{\nu\alpha}\Delta\mathscr{C}^{(0)\beta}\right)
+ \frac{1}{3} \epsilon^{\nu\alpha\beta\rho}\Delta \mathscr{C}^{(0)}_{5,\rho},
\end{eqnarray}
because these new collision functions with tilde  always appear as a whole.

The Eqs.(\ref{Js-cs-t-2-n}), (\ref{S-cs-s-2-n}),(\ref{F-eq-2-bar}) and (\ref{S-cs-t-2-n}) expresses other second-order
 Wigner functions in terms of ${\mathscr{J}}_{s,n}^{(2)}$, ${\mathscr{M}}^{(2)}_\mu$,
\begin{eqnarray}
\label{Jbar-2}
\bar{\mathscr{J}}_{s,\mu}^{(2)}
&=& \frac{\bar p_\mu}{p_n} {\mathscr{J}}_{s,n}^{(2)}-\frac{s m}{2p_n}{\mathscr{M}}^{(2)}_\mu
+\frac{s}{2p_n}\bar\epsilon_{\mu\alpha\beta}\partial_x^{\alpha}{\mathscr{J}}^{(1)\beta}_s
+\frac{s}{2p_n}\bar\epsilon_{\mu\alpha\beta}\tilde{ \mathscr{C}}^{(1)\alpha\beta}_{s},\\
\label{F-2}
{\mathscr{F}}^{(2)}   &=&  \frac{ m}{p_n} \sum_s{\mathscr{J}}_{s,n}^{(2)}
-\frac{ 1}{2p_n}  n^\mu\partial_x^{\nu}{\mathscr{S}}_{\mu\nu}^{(1)}
-\frac{ 1}{2p_n} n^\mu {\tilde{\mathscr{C}}^{(1)\nu}}_{\ \ \ \ \ \mu\nu},
\\
\label{K-2}
\mathscr{K}^{(2)\mu}  &=& \frac{1}{p_n}\bar\epsilon^{\mu\nu\alpha}\bar p_\nu \mathscr{M}_\alpha^{(2)}
+\frac{ 1}{2p_n}\Delta^{\mu\lambda}\partial_\lambda^x {\mathscr{F}}^{(1)}
+\frac{ 1}{2p_n} \Delta^{\mu\lambda}\tilde{\mathscr{C}}^{(1)}_\lambda,\\
\label{P-2}
{\mathscr{P}}^{(2)} &=& \frac{1}{4p_n}\bar \epsilon_{\nu\alpha\beta}\partial^{\nu}_x {\mathscr{S}}^{(1)\alpha\beta}
+\frac{1}{4p_n}\bar \epsilon_{\nu\alpha\beta}\tilde{\mathscr{C}}^{(1)\nu\alpha\beta}.
\end{eqnarray}
The antisymmetric tensor Wigner function follows as
\begin{eqnarray}
\label{S-M-2}
\mathscr{S}^{(2)}_{\mu\nu}&=&\frac{1}{p_n} \epsilon_{\mu\nu\alpha\beta}p^\alpha \mathscr{M}^{(2)\beta}
+\frac{ 1}{2p_n}\left(\Delta_\mu^\lambda n_\nu - \Delta_\nu^\lambda n_\mu\right)
\left(\partial_\lambda^x {\mathscr{F}}^{(1)} + \tilde{\mathscr{C}}^{(1)}_{\lambda}\right).
\end{eqnarray}
Substituting the expression (\ref{P-2}) into  Eqs.(\ref{Js-eq-2-nbar}) and (\ref{S-cs-t-2-bar}), we obtain  the kinetic equation for ${\mathscr{J}}_{s,n}^{(1)}$ and ${\mathscr{M}}^{(1)}_\mu$,
respectively,
\begin{eqnarray}
\label{Jsn-eq-1}
p^\mu \partial^x_\mu\left( \frac{\mathscr{J}_{s,n}^{(1)}}{p_n}\right)
&=&-\frac{s}{2p_n^2}  (\partial^x_\nu n_\mu)\left[ p^\nu \left(m {\mathscr{M}}^{(1)\mu}
-\bar\epsilon^{\mu\rho\sigma} \partial^x_\rho  \mathscr{J}_{s,\sigma}^{(0)} \right)
+ \bar\epsilon^{\mu\nu\beta}p^\lambda
\left(\mathscr{C}^{(0)}_{s,\beta\lambda}-\mathscr{C}^{(0)}_{s,\lambda\beta}\right)\right]
\nonumber\\
& &-\tilde{\mathscr{C}}^{(1)\mu}_{s\mu}-\frac{ms}{4p_n}\bar \epsilon_{\nu\alpha\beta}\tilde{\mathscr{C}}^{(1)\nu\alpha\beta}
{-}\frac{s}{2}\partial^\mu_x \left(\frac{1}{p_n}\bar\epsilon_{\mu\alpha\beta}
\mathscr{C}^{(0)\alpha\beta}_{s}\right),
\\
\label{M-eq-1}
p^\nu\partial_{\nu}^x\left(\frac{\mathscr{M}^{(1)\mu}}{p_n}\right)
&=&-\frac{1}{p_n^2}   (\partial^x_\lambda n_\nu) p^\lambda \left(p^\mu \mathscr{M}^{(1)\nu}
- \frac{1}{2} \bar\epsilon^{\mu\nu\rho}\partial^x_\rho\mathscr{F}^{(0)}
 -\frac{1}{2} \bar\epsilon^{\mu\nu\lambda} p^\sigma\mathscr{C}^{(0)}_\sigma\right)
\nonumber\\
& &-\frac{1}{2}\left(\epsilon^{\mu\nu\alpha\beta} + \frac{p^\mu}{p_n}\bar\epsilon^{\nu\alpha\beta}\right)
\left[\tilde{\mathscr{C}}^{(1)}_{\nu\alpha\beta}
+ \partial^x_\nu
\left(\frac{\bar{\mathscr{C}}^{(0)}_\alpha n_\beta - \bar{\mathscr{C}}^{(0)}_\beta n_\alpha}{2p_n}\right)\right].
\end{eqnarray}
The equation (\ref{P-eq-2-n}) gives longitudinal  constraint condition for ${\mathscr{M}}^{(2)}_\mu$
\begin{eqnarray}
\label{M-cs-2}
\bar p_\mu {\mathscr{M}}^{(2)\mu}  &=& m \sum_s s {\mathscr{J}}_{s,n}^{(2)}
- \frac{1}{2}\tilde{\mathscr{C}}^{(1)}_{5, n}
-\frac{1}{2}n^\mu \partial^x_\mu {\mathscr{P}}^{(1)}.
\end{eqnarray}
From Eq.(\ref{Js-cs-s-2-nbar}) and Eq.(\ref{P-eq-2-bar}), we obtain the modification to the  on-shell conditions of ${\mathscr{J}}_{s,n}^{(2)}$ and ${\mathscr{M}}^{(2)}_\mu$, respectively,
\begin{eqnarray}
\label{Jn-os-2}
& &\left(p^2-m^2\right)\frac{ {\mathscr{J}}_{s,n}^{(2)}}{p_n}\nonumber\\
&=&-\frac{m}{8p_n} n^\mu\partial_x^{\nu}
\left[\frac{\Delta_\mu^\lambda n_\nu - \Delta_\nu^\lambda n_\mu}{p_n}\left(\partial_\lambda^x {\mathscr{F}}^{(0)}
+\mathscr{C}^{(0)}_{\lambda}\right)\right]
-\frac{m}{4p_n}  n^\mu{\tilde{\mathscr{C}}^{(1)\nu}}_{\ \ \ \ \ \mu\nu}
\nonumber\\
& &-\frac{s m}{4p_n}\left(\tilde{\mathscr{C}}^{(1)}_{5, n}
+ n^\mu \partial^x_\mu {\mathscr{P}}^{(1)}\right)
-\frac{1}{4p_n}\bar\epsilon_{\mu\alpha\beta}p^\mu \partial_x^{\alpha}
\left[\frac{\bar\epsilon^{\beta\rho\sigma}}{p_n}\left(\partial^x_{\rho}{\mathscr{J}}^{(0)}_{s,\sigma}
+ \mathscr{C}^{(0)}_{s,\rho\sigma}\right) \right]
\nonumber\\
& &-\frac{s}{2p_n}\bar\epsilon_{\mu\alpha\beta}p^\mu\tilde{\mathscr{C}}^{(1)\alpha\beta}_{s}
-\frac{1}{2}\Delta\mathscr{C}_{s,\mu}^{(0)\mu},
\end{eqnarray}
\begin{eqnarray}
\label{M-os-2}
& &\left(p^2-m^2\right)\frac{ \mathscr{M}^{(2)}_\mu}{p_n}\nonumber\\
&=&\frac{1}{2}\Delta_\mu^\lambda \partial^x_\lambda {\mathscr{P}}^{(1)}
+\frac{1}{2}\Delta^\lambda_\mu \tilde{\mathscr{C}}^{(1)}_{5,\lambda}
-\frac{m}{4p_n} \sum_s \bar\epsilon_{\mu\alpha\beta}\partial_x^{\alpha}
\left[\frac{s\bar\epsilon^{\beta\rho\sigma}}{p_n}\left(\partial^x_{\rho}{\mathscr{J}}^{(0)}_{s,\sigma}
+ \mathscr{C}^{(0)}_{s,\rho\sigma}\right) \right]
\nonumber\\
& &-\frac{m}{2p_n} \bar\epsilon_{\mu\alpha\beta}  \sum_s \tilde{\mathscr{C}}^{(1)\alpha\beta}_{s}
-\frac{\bar p_\mu}{2 p_n} \left(\tilde{\mathscr{C}}^{(1)}_{5, n}
+n^\nu \partial^x_\nu {\mathscr{P}}^{(1)}\right)\nonumber\\
& &+\frac{1}{4p_n}\bar\epsilon_{\mu\nu\alpha}\bar p^\nu
\Delta^{\alpha\lambda}\partial_\lambda^x
\left[\frac{ n^\mu}{p_n}\left(  \partial_x^{\nu}{\mathscr{S}}_{\mu\nu}^{(0)}
+{\mathscr{C}^{(0)\nu}}_{\mu\nu}\right)\right]
-\frac{1}{2p_n}\bar\epsilon_{\mu\nu\alpha}\bar p^\nu  \tilde{\mathscr{C}}^{(1)\alpha}.
\end{eqnarray}

Following the same procedure as we have taken at first order, we can obtain the self-consistent constraints for the collision terms at the first order.
Besides much more complicated vector algebraic and  derivative operation at second order,   we also encounter the terms associated with the derivative terms with  $n^\mu$ such as
$\partial_x^\nu n^\mu$ in the second-order constraint equations. However, it is remarkable that all these terms cancel each other if we impose  the zeroth-order constraint equations (\ref{collision-0-f})-(\ref{collision-4-f}). The final independent constraint equations are given by
\begin{eqnarray}
\label{collision-0-f-2}
p^\mu \tilde{\mathscr{C}}^{(1)}_{5,\mu}
&=&\frac{1}{4} \epsilon^{\mu\nu\alpha\beta}\partial_{\mu}^x\mathscr{C}^{(0)}_{\nu\alpha\beta}
-m\sum_s s\Delta\mathscr{C}_{s,\mu}^{(0)\mu},
\\
\label{collision-1-f-2}
m\sum_s\tilde{\mathscr{C}}^{(1)\lambda}_{s,\lambda}&=&p^\lambda \tilde{\mathscr{C}}^{(1)}_{\lambda}
-\frac{ 1}{2}\partial_x^\nu{\mathscr{C}^{(0)\lambda}}_{\nu\lambda},
\\
\label{collision-2-f-2}
2mp_\mu\sum_s s\tilde{\mathscr{C}}^{(1)\lambda}_{s,\lambda}
&=&[p_\mu p^\lambda \epsilon_{\lambda\nu\alpha\beta} - (p^2-m^2)\epsilon_{\mu\nu\alpha\beta}]
\tilde{\mathscr{C}}^{(1)\nu\alpha\beta}
\nonumber\\
& &+\epsilon_{\mu\lambda\alpha\beta}\partial^\lambda_x  (m\sum_s  \mathscr{C}^{(0)\alpha\beta}_{s}
-p^\beta \mathscr{C}^{(0)\alpha})
- p^\nu\partial_\nu^x\mathscr{C}^{(0)}_{5, \mu},
\\
\label{collision-3-f-2}
m\epsilon_{\mu\nu\alpha\beta} \sum_s s\tilde{\mathscr{C}}^{(1)\alpha\beta}_{s}
&=& p_\nu{\tilde{\mathscr{C}}^{(1)\lambda}}_{\ \ \ \ \ \mu\lambda}
-p_\mu{\tilde{\mathscr{C}}^{(1)\lambda}}_{\ \ \ \ \ \nu\lambda}
-\frac{1}{2}\epsilon_{\mu\nu\rho\sigma}p^\rho \epsilon^{\sigma\lambda\alpha\beta}\tilde{\mathscr{C}}^{(1)}_{\lambda\alpha\beta}
\nonumber\\
& &+\frac{ 1}{2} (\partial^x_{\nu}\mathscr{C}^{(0)}_\mu-\partial^x_{\mu}\mathscr{C}^{(0)}_\nu)
-\frac{1}{2}\epsilon_{\mu\nu\rho\sigma}\partial_x^{\rho}{ \mathscr{C}}^{(0)\sigma}_{5},
\\
\label{collision-4-f-2}
\frac{ m }{2} \tilde{\mathscr{C}}^{(1)\mu}
+\frac{sm}{4}\epsilon^{\mu\beta\rho\sigma}\tilde{\mathscr{C}}^{(1)}_{\beta\rho\sigma}
&=&p_\lambda (\tilde{\mathscr{C}}^{(1)\mu\lambda}_{s}
-\tilde{\mathscr{C}}^{(1)\lambda\mu}_{s})
+p^\mu \tilde{\mathscr{C}}^{(1)\lambda}_{s,\lambda}
-\frac{s}{2}\epsilon^{\mu\beta\rho\sigma}\partial_\beta^x \mathscr{C}^{(0)}_{s,\rho\sigma}.
\end{eqnarray}

\section{Some other constraint conditions}

In addition to the  constraint equations  for zeroth order (\ref{collision-0-f})-(\ref{collision-4-f})  and first order (\ref{collision-0-f-2})-(\ref{collision-4-f-2}) 
which are derived from the self-consistency of the Wigner equations,  we can obtain other  constraint equations from the specific structure defined from (\ref{def1})
to (\ref{def8}). The antisymmetry of the electromagnetic field tensor $F^{\mu\lambda}$ requires the following constraints
\begin{eqnarray}
\label{dC}
\partial_\nu^p {\mathscr{C}^{\nu}} &=& 0,\ \ \ 
\partial_\nu^p {\mathscr{C}^{\nu}_5} = 0,\ \ \ 
\partial_\mu^p \mathscr{C}^{\mu\nu}_s =0, \ \ \ 
\partial_\nu^p {\mathscr{C}^{\nu\alpha\beta}}=0, \\
\label{dDC}
\partial_\nu^p \Delta{\mathscr{C}^{\nu}} &=& 0,\ \
\partial_\nu^p \Delta{\mathscr{C}^{\nu}_5} = 0,\ \
\partial_\mu^p \Delta\mathscr{C}^{\mu\nu}_s =0, \ \
\partial_\nu^p \Delta{\mathscr{C}^{\nu\alpha\beta}}=0.
\end{eqnarray}
The constraints (\ref{dC}) require that the zeroth-order ${\mathscr{C}^{(0)\nu}}$, $ {\mathscr{C}^{(0)\nu}_5} $, $\mathscr{C}^{(0)\mu\nu}_s$ and ${\mathscr{C}^{(0)\nu\alpha\beta}}$ 
in (\ref{collision-0-f})-(\ref{collision-4-f}) should satisfy the same constraints. However, the constraints (\ref{dC}) and (\ref{dDC}) cannot  lead to the similar constraints on  the first-order ${\tilde{\mathscr{C}}^{(1)\nu}}$, $ {\tilde{\mathscr{C}}^{(1)\nu}_5} $, $\tilde{\mathscr{C}}^{(1)\mu\nu}_s$ and $\tilde{\mathscr{C}}^{(1)\nu\alpha\beta}$  in (\ref{dC}) and (\ref{dDC}) due to the out-of-step linear combination in the definitions (\ref{tC-1})-(\ref{tC-4}). In the following sections, we will disregard these constraints and only focus on the 
constraints (\ref{collision-0-f})-(\ref{collision-4-f})  and  (\ref{collision-0-f-2})-(\ref{collision-4-f-2}).

\section{Self-consistent specific  solutions for collision terms}
\label{sec:solution}
As we all know,  these collision terms cannot be  determined by a group of closed  equations due to  BBGKY  hierarchy. Some proper approximation must be imposed to make the equations closed.
No matter what approximation we make, the self-consistent constraints  (\ref{collision-0-f})-(\ref{collision-4-f})  and  (\ref{collision-0-f-2})-(\ref{collision-4-f-2}) derived in  previous sections should be fulfilled.
In this section, we will find  self-consistent particular  solutions for these collision terms.

Let us start from the zeroth-order constraints conditions (\ref{collision-0-f})-(\ref{collision-4-f}) and try to find a specific expression as simple as possible but still not very trivial.
In the simplest case, we will assume that all the collision terms at the first order are on-shell. It is easy to verify that the following expressions always satisfy all  first-order constraints conditions (\ref{collision-0-f})-(\ref{collision-4-f}),
\begin{eqnarray}
\mathscr{C}_{5,\mu}^{(0)}&=&0,
\\
\mathscr{C}^{(0)\mu}&=& m \sum_s \mathscr{X}_s^{(0)\mu}  ,
\\
\mathscr{C}^{(0)\mu\nu}_{s}&=& \mathscr{X}_s^{(0)\mu} p^\nu,
\\
\mathscr{C}^{(0)\nu\alpha\beta}&=&-\frac{m}{3}\epsilon^{\mu\nu\alpha\beta}\sum_s s\mathscr{X}^{(0)}_{s,\mu},
\end{eqnarray}
where  $\mathscr{X}_s^{(0)\mu} $ with chirality index $s$ can be arbitrary vector function with the on-shell condition,
\begin{eqnarray}
\label{Y-os-0}
\left(p^2-m^2\right) {\mathscr{X}}_{s}^{(0)} = 0,\ \ \ \textrm{or equivalently, } \ \ \
 {\mathscr{X}}_{s}^{(0)} = {\mathcal{X}}_{s}^{(0)}\delta(p^2-m^2).
\end{eqnarray}
The physical meaning of function $\mathscr{X}_s^{(0)\mu} $
depends on the specific system we are considering. In the chiral limit $m=0$, we find that only $\mathscr{C}^{(0)\mu\nu}_{s}$ survive and
$\mathscr{C}^{(0)\mu}$ and $\mathscr{C}^{(0)\nu\alpha\beta}$  both vanish and are trivial. In order to make $\mathscr{C}^{(0)\nu\alpha\beta}$
not trivial in the chiral limit because it is related to the magnetic moment distribution, we can find a slightly more complex solution
\begin{eqnarray}
\mathscr{C}_{5\mu}^{(0)}&=&0,
\\
\mathscr{C}^{(0)\mu}&=& m \sum_s \mathscr{X}_s^{(0)\mu} ,
\\
\mathscr{C}^{(0)\mu\nu}_{s}&=& \mathscr{X}_s^{(0)\mu} p^\nu
+ \frac{s m}{4 u\cdot p} \left(u^\mu \mathscr{Y}^{(0)\nu} - u^\nu\mathscr{Y}^{(0)\mu} \right) ,
\\
\mathscr{C}^{(0)\nu\alpha\beta}&=&-\frac{u^\nu}{u\cdot p}\epsilon^{\alpha\beta\rho\sigma}p_\rho\mathscr{Y}^{(0)}_\sigma
-\frac{m}{3}\epsilon^{\mu\nu\alpha\beta}\sum_s s\mathscr{X}^{(0)}_{s,\mu},
\end{eqnarray}
where we have introduced the vector functions $u^\mu$ and  $\mathscr{Y}^{(0)\mu}$ which satisfy
\begin{eqnarray}
\label{X-con}
u^2=1,\ \ \ u\cdot\mathscr{Y}^{(0)}=0,\ \ \ p\cdot\mathscr{Y}^{(0)}=0,
\end{eqnarray}
and the on-shell condition
\begin{eqnarray}
\label{X-os-0}
\left(p^2-m^2\right) {\mathscr{Y}}^{(0)} = 0,\ \ \ \textrm{or equivalently, } \ \ \
 {\mathscr{Y}}^{(0)} = {\mathcal{Y}}^{(0)}\delta(p^2-m^2).
\end{eqnarray}
In the chiral limit, we have
\begin{eqnarray}
\mathscr{C}_{5,\mu}^{(0)}&=&0,
\\
\mathscr{C}^{(0)\mu}&=& 0 ,
\\
\mathscr{C}^{(0)\mu\nu}_{s}&=&\mathscr{X}_s^{(0)\mu} p^\nu,
\\
\mathscr{C}^{(0)\nu\alpha\beta}&=&-\frac{1}{u\cdot p}u^\nu\epsilon^{\alpha\beta\rho\sigma}p_\rho\mathscr{Y}^{(0)}_\sigma.
\end{eqnarray}
It should be pointed out  that the normalized time-like vector $u^\mu$ should be regarded as a physical function which could depend on both coordinate and momentum and has no relations to  $n^\mu$.
With this specific expressions, we find that these collision terms do not  modify either  longitudinal constraint condition for ${\mathscr{M}}^{(1)}_\mu$
\begin{eqnarray}
\label{M-cs-1-RTA-2}
\bar p_\mu {\mathscr{M}}^{(1)\mu}  &=& m \sum_s s {\mathscr{J}}_{s,n}^{(1)},
\end{eqnarray}
or  the on-shell conditions of ${\mathscr{J}}_{s,n}^{(1)}$ and ${\mathscr{M}}^{(1)}_\mu$
\begin{eqnarray}
\label{Jn-os-1-RTA-2}
\left(p^2-m^2\right)\frac{ {\mathscr{J}}_{s,n}^{(1)}}{p_n}
&=&0,
\\
\label{M-os-1-RTA-2}
\left(p^2-m^2\right)\frac{ {\mathscr{M}}_{\mu}^{(1)}}{p_n}
&=& 0.
\end{eqnarray}
The general expressions are given by
\begin{eqnarray}
\label{Jn-os-0-r}
 {\mathscr{J}}_{s,n}^{(1)} &=& p_n {\mathcal{J}}_{s,n}^{(1)}\delta(p^2-m^2) ,\\
\label{M-os-0-r}
 {\mathscr{M}}_{\mu}^{(1)} &=& p_n {\mathcal{M}}_{\mu}^{(1)}\delta(p^2-m^2).
\end{eqnarray}
The quantum  kinetic equations of ${\mathscr{J}}_{s,n}^{(0)}$ and ${\mathscr{M}}^{(0)}_\mu$ with these collision terms follows
\begin{eqnarray}
\label{Jsn-eq-0-RTA-2}
p^\mu \partial^x_\mu\left( \frac{\mathscr{J}_{sn}^{(0)}}{p_n}\right)
&=&-\frac{ m s}{2p_n}p^\nu \left(\partial^x_\nu  n_\mu \right)\frac{{\mathscr{M}}^{(0)\mu} }{p_n}\nonumber\\
& &-\mathscr{X}^{(0)\mu}_{s}p_\mu
+\frac{s m^2}{2p_n}\sum_{s'}s'\mathscr{X}^{(0)n}_{s'}
-\frac{ms}{2p_n} \mathscr{Y}_n^{(0)},  \\
\label{M-eq-0-RTA-2}
p^\nu \partial^x_\nu \left(\frac{\mathscr{M}^{(0)\mu}}{p_n}\right)
&=&-\frac{1}{p_n} p^\mu p^\nu \left( \partial^x_\nu n_\lambda \right)\frac{\mathscr{M}^{(0)\lambda}}{p_n}\nonumber\\
& & -\frac{1}{p_n}\left( p^\mu\mathscr{Y}_n^{(0)} - p_n \mathscr{Y}^{(0)\mu}\right)
-m\sum_{s'}s' \left(\mathscr{X}^{(0)\mu}_{s'}-\frac{p^\mu}{p_n}\mathscr{X}^{(0)n}_{s'} \right).
\end{eqnarray}
From the second-order constraints (\ref{collision-0-f-2})-(\ref{collision-4-f-2}), we can find a first-order solution
\begin{eqnarray}
\tilde{\mathscr{C}}_{5,\mu}^{(1)}&=& \frac{u_\mu}{2 u\cdot p} \left(-\partial^\nu_x\mathscr{Y}^{(0)}_\nu
+ m \sum_s s\partial^\nu_x \mathscr{X}^{(0)}_{s,\nu}\right),
\\
\tilde{\mathscr{C}}^{(1)\mu}&=& m \sum_s \mathscr{X}_s^{(1)\mu}
-\epsilon^{\mu\nu\alpha\beta} \partial^x_\nu \left(\frac{ u_\alpha \mathscr{Y}_\beta^{(0)} }{2 u\cdot p}\right),
\\
\tilde{\mathscr{C}}^{(1)\mu\nu}_{s}&=& \mathscr{X}_s^{(1)\mu} p^\nu
+ \frac{s m}{4 u\cdot p} \left(u^\mu\mathscr{Y}^{(1)\nu} - u^\nu\mathscr{Y}^{(1)\mu}\right) ,
\\
\tilde{\mathscr{C}}^{(1)\nu\alpha\beta}&=&-\frac{1}{u\cdot p}u^\nu\epsilon^{\alpha\beta\rho\sigma}p_\rho\mathscr{Y}^{(1)}_\sigma
-\frac{m}{3}\epsilon^{\mu\nu\alpha\beta}\sum_s s\mathscr{X}^{(1)}_{s,\mu}.
\end{eqnarray}
With these specific expressions,  first-order Wigner functions have the following form
\begin{eqnarray}
{\mathscr{J}}_{s,\mu}^{(1)}
&=& \frac{ p_\mu}{p_n} {\mathscr{J}}_{s,n}^{(1)}-\frac{s m}{2p_n}{\mathscr{M}}^{(1)}_\mu
+\frac{s}{2p_n}\bar\epsilon_{\mu\alpha\beta}\partial_x^{\alpha}{\mathscr{J}}^{(0)\beta}_s
\nonumber\\
& &+\frac{1}{2p_n}\bar\epsilon_{\mu\alpha\beta}\left(s\mathscr{X}_s^{(0)\alpha} p^\beta
+ \frac{m}{2u\cdot p}u^\alpha\mathscr{Y}^{(0)\beta}\right),
\\
{\mathscr{F}}^{(1)}   &=&  \frac{ m}{p_n} \sum_s{\mathscr{J}}_{s,n}^{(1)}
-\frac{ 1}{2p_n}  n^\mu\partial_x^{\nu}{\mathscr{S}}_{\mu\nu}^{(0)}
-\frac{ 1}{2p_n(u\cdot p)} u_\nu\bar\epsilon^{\nu\rho\sigma} p_\rho\mathscr{Y}^{(0)}_\sigma,
\\
{\mathscr{P}}^{(1)} &=& \frac{1}{4p_n}\bar \epsilon_{\nu\alpha\beta}\partial^{\nu}_x {\mathscr{S}}^{(0)\alpha\beta}
+\frac{1}{2p_n}\left(\mathscr{Y}^{(0)}_n -m\sum_s s\mathscr{X}_{sn}^{(0)}  \right),
\\
\mathscr{K}^{(1)\mu}  &=& \frac{1}{p_n}\bar\epsilon^{\mu\nu\alpha}\bar p_\nu \mathscr{M}_\alpha^{(1)}
+\frac{ 1}{2p_n}\Delta^{\mu\lambda}\partial_\lambda^x {\mathscr{F}}^{(0)}
+\frac{ m}{2p_n}\sum_s \bar{\mathscr{X}}_s^{(0)\mu}.
\end{eqnarray}
The quantum kinetic equations of ${\mathscr{J}}_{s,n}^{(1)}$ and ${\mathscr{M}}^{(1)}_\mu$ are given by
\begin{eqnarray}
\label{Jsn-eq-1-RTA-2}
p^\mu \partial^x_\mu\left( \frac{\mathscr{J}_{s,n}^{(1)}}{p_n}\right)
&=&-\frac{s}{2p_n} p^\nu (\partial^x_\nu n_\mu)\left(  m \frac{ {\mathscr{M}}^{(1)\mu}}{p_n}
- \frac{1}{p_n}\bar\epsilon^{\mu\alpha\beta} \partial^x_\alpha  \mathscr{J}_{s,\beta}^{(0)} \right)\nonumber\\
& &+\frac{s}{2p_n^2} p^\nu (\partial^x_\nu n_\mu)\bar\epsilon^{\mu\alpha\beta}p_\beta  \mathscr{X}_{s,\alpha}^{(0)}
 -\frac{s}{2p_n}\bar\epsilon^{\mu\alpha\beta}p_\beta \partial^x_\mu  \mathscr{X}_{s,\alpha}^{(0)} \nonumber\\
& &-\left(\mathscr{X}^{(1)\mu}_{s}p_\mu
-\frac{s m^2}{2p_n}\sum_{s'}s'\mathscr{X}^{(1)n}_{s'}\right)
-\frac{ms}{2p_n}\mathscr{Y}_n^{(1)},\\
\label{M-eq-1-RTA-2}
p^\nu \partial^x_\nu \left(\frac{\mathscr{M}^{(1)\mu}}{p_n}\right)
&=&-\frac{1}{p_n}  p^\lambda   (\partial_{\lambda}^x n_\nu)
\left( p^\mu \frac{\mathscr{M}^{(1)\nu} }{p_n}
-\frac{1}{2p_n}\bar\epsilon^{\mu\nu\alpha}
\partial_\alpha^x {\mathscr{F}}^{(0)}\right)\nonumber\\
& &+\frac{m}{2p_n^2}  p^\lambda   (\partial_{\lambda}^x n_\nu)
\bar\epsilon^{\mu\nu\alpha} \sum_{s'} \mathscr{X}^{(0)}_{s',\alpha}
-\frac{m}{2p_n}  \bar\epsilon^{\mu\nu\alpha}\partial_{\nu}^x \sum_{s'} \mathscr{X}^{(0)}_{s',\alpha}  \nonumber\\
& & -m\sum_{s'}s' \left(\mathscr{X}^{(1)\mu}_{s'}-\frac{p^\mu}{p_n}\mathscr{X}^{(1)n}_{s'} \right)
-\frac{1}{p_n}\left( p^\mu \mathscr{Y}_n^{(1)} - p_n\mathscr{Y}^{(1)\mu}\right).
\end{eqnarray}

\section{Self-consistent relaxation-time approximation}
\label{sec:RTA}
The general kinetic equation is very difficult to tackle because the BBGKY   hierarchy or the non-linear collision terms.
The relaxation-time approximation  has been proposed since 1950s \cite{Bhatnagar:1954zz,Marle:1969,Anderson:1974nyl} and used quite successfully in several field physics. Recently, the quantum kinetic equation at naive relaxation-time approximation has been discussed in \cite{Hidaka:2017auj,Wang:2021qnt}.
In this section, we will present the quantum kinetic equation at self-consistent relaxation-time approximation, which is consistent with the self-consistent constraints obtained in previous sections.

From the requirement (\ref{X-con}), we can assume the collision terms $\mathscr{X}_{s\mu}^{(k)}$ and $\mathscr{Y}_{\mu}^{(k)}$  ($k=0,1$) take the conventional  form  at relaxation-time approximation
\begin{eqnarray}
\mathscr{X}_{s\mu}^{(k)}=\frac{1}{p_u}\left(\frac{ u_\mu}{\tau_{1s}}+\frac{ p_\mu}{p_u\tau_{2s}}\right)\delta{\mathscr{J}}_{s,u}^{(k)}
+\frac{ m s}{2 p_u^2\tau_{3}}\delta{\mathscr{M}}_{u\perp,\mu}^{(k)} ,\ \ \
\mathscr{Y}_{\mu}^{(k)}= -\frac{1}{\tau_{4}}\delta{\mathscr{M}}_{u\perp,\mu}^{(k)},
\end{eqnarray}
where $\tau_{1s}$, $\tau_{2s}$ , $\tau_3$, and $\tau_4$ denote the relaxation time parameters associtated with distribution functions
${\mathscr{J}}_{s,n}$ and ${\mathscr{M}}_{\perp,\mu}$. In this section, we identify the time-like vector $u^\mu$ as the fluid velocity.
 We have decomposed the functions along   $u^\mu$ instead of $n^\mu$  because all the collision functions cannot  depends on $n^\mu$  and they can only depend on some physical quantity such as $u^\mu$. The symbols in the above expressions are defined by
\begin{eqnarray}
& & p_u = u\cdot p,\ \ \ \  {\mathscr{J}}_{s,u}^{(k)} = u\cdot {\mathscr{J}}_{s}^{(k)},\ \ \
{\mathscr{M}}_{u\perp,\mu}^{(k)} = \frac{1}{2}\epsilon_{\mu\nu\alpha\beta}u^\nu \mathscr{S}^{(k)\alpha\beta}, \\
& & \delta \mathscr{J}_{s,u}^{(k)} ={\mathscr{J}}_{s,u}^{(k)} - {\mathscr{J}}_{s,u,\textrm{eq}}^{(k)},\ \ \
\delta{\mathscr{M}}_{u\perp,\mu}^{(k)} = {\mathscr{M}}_{u\perp,\mu}^{(k)} - {\mathscr{M}}_{u\perp,\textrm{eq},\mu}^{(k)},
\end{eqnarray}
where the subscript `eq' indicates  the corresponding functions at local or global equilibrium.
With these expressions, the quantum kinetic equations   are given by
\begin{eqnarray}
\label{Jsn-eq-0-RTA-2}
p^\mu \partial^x_\mu\left( \frac{\mathscr{J}_{sn}^{(k)}}{p_n}\right)
&=&-\frac{  s}{2p_n}p^\nu \left(\partial^x_\nu  n_\mu \right)
\left(  m \frac{ {\mathscr{M}}^{(k)\mu}}{p_n}
- \frac{1}{p_n}\bar\epsilon^{\mu\alpha\beta} \partial^x_\alpha  \mathscr{J}_{s\beta}^{(k-1)} \right)\nonumber\\
& &- \frac{p_\mu}{p_u}\left( \frac{u^\mu}{\tau_{1s}}+\frac{p^\mu}{p_u\tau_{2s}}\right)
\delta{\mathscr{J}}_{s,u}^{(k)}
+\frac{sm^2}{2p_n p_u}\sum_{s'} s' \left( \frac{n\cdot u}{\tau_{1s'}}+\frac{p_n}{p_u\tau_{2s'}}\right)
\delta{\mathscr{J}}_{s',u}^{(k)}\nonumber\\
& &+\frac{s}{2p_n^2}\bar\epsilon^{\mu\alpha\beta}p_\beta \left[ p^\nu (\partial^x_\nu n_\mu)  - p_n  \partial^x_\mu \right]
\left(\frac{ u_\alpha}{p_u\tau_{1s}}\delta{\mathscr{J}}_{s,u}^{(k-1)}\right)\nonumber\\
& &+\frac{ms }{2p_n p_u}\left(\frac{p_u}{\tau_4} + \frac{m^2}{p_u \tau_3}\right)n^\mu
\delta{\mathscr{M}}_{u\perp,\mu}^{(k)} \nonumber\\
& &+\frac{m}{4p_n^2}\bar\epsilon^{\mu\alpha\beta}p_\beta \left[ p^\nu (\partial^x_\nu n_\mu)  - p_n  \partial^x_\mu \right]
\left(\frac{1}{p_u^2\tau_{3}} {\mathscr{M}}_{u\perp,\alpha}^{(k-1)}\right),
\end{eqnarray}

\begin{eqnarray}
\label{M-eq-0-RTA-2}
p^\nu \partial^x_\nu \left(\frac{\mathscr{M}^{(k)\mu}}{p_n}\right)
&=&-\frac{1}{p_n}  p^\lambda   (\partial_{\lambda}^x n_\nu)
\left( p^\mu \frac{\mathscr{M}^{(k)\nu} }{p_n}
-\frac{1}{2p_n}\bar\epsilon^{\mu\nu\alpha}
\partial_\alpha^x {\mathscr{F}}^{(k-1)}\right)\nonumber\\
& & +\frac{ p^\mu n_\lambda - p_n g^\mu_\lambda}{p_n p_u}\left(\frac{p_u}{\tau_4} + \frac{m^2}{p_u \tau_3}\right)
\delta{\mathscr{M}}_{u\perp}^{(k) \lambda}\nonumber\\
& &+\frac{m \left(p^\mu n_\lambda -p_n g^{\mu}_\lambda\right)}{p_n p_u}\sum_{s'} s' \left( \frac{u^\lambda}{\tau_{1s'}}+\frac{p^\lambda}{p_u\tau_{2s'}}\right)
\delta{\mathscr{J}}_{s',u}^{(k)}\nonumber\\
& &+\frac{m}{2p_n^2}\bar\epsilon^{\mu\nu\alpha}\left[  p^\lambda   (\partial_{\lambda}^x n_\nu)
-p_n \partial_{\nu}^x\right]
\sum_{s'}\frac{1}{p_u} \left(\frac{ u_\alpha}{\tau_{1s'}}+\frac{ p_\alpha}{p_u\tau_{2s'}}\right)
\delta{\mathscr{J}}_{s',u}^{(k-1)}.
\end{eqnarray}
It should be noted that when $k=0$ we define the functions with superscript $k=-1$ to vanish.
The zeroth-order Wigner functions include no collision contribution. The Wigner functions at first order  read
\begin{eqnarray}
{\mathscr{J}}_{s,\mu}^{(1)}
&=& \frac{ p_\mu}{p_n} {\mathscr{J}}_{s,n}^{(1)}-\frac{s m}{2p_n}{\mathscr{M}}^{(1)}_\mu
+\frac{s}{2p_n}\bar\epsilon_{\mu\alpha\beta}\partial_x^{\alpha}{\mathscr{J}}^{(0)\beta}_s
\nonumber\\
& &+\frac{s}{2p_n p_u \tau_{1s}}\bar\epsilon_{\mu\alpha\beta}u^\alpha p^\beta
\delta{\mathscr{J}}_{s',u}^{(0)}
+\frac{m}{4p_n p_u^2}\bar\epsilon_{\mu\alpha\beta}
\left( \frac{p^\beta}{\tau_3}  + \frac{p_u  u^\beta}{\tau_4} \right)
\delta{\mathscr{M}}_{u\perp}^{(0) \alpha},\\
{\mathscr{F}}^{(1)}   &=&  \frac{ m}{p_n} \sum_s{\mathscr{J}}_{s,n}^{(1)}
-\frac{ 1}{2p_n}  n^\mu\partial_x^{\nu}{\mathscr{S}}_{\mu\nu}^{(0)}
+\frac{ 1}{2p_n p_u \tau_4} u_\nu\bar\epsilon^{\nu\rho\sigma} p_\rho
\delta{\mathscr{M}}_{u\perp,\sigma}^{(0)},\\
{\mathscr{P}}^{(1)} &=& \frac{1}{4p_n}\bar \epsilon_{\nu\alpha\beta}\partial^{\nu}_x {\mathscr{S}}^{(0)\alpha\beta}
-\frac{m}{2p_n p_u}\sum_s s  \left(\frac{ n\cdot u}{\tau_{1s}}+\frac{ p_n}{p_u\tau_{2s}}\right)
\delta{\mathscr{J}}_{s,u}^{(0)} \nonumber\\
& &-\frac{1}{2p_n}\left(\frac{1}{\tau_4} + \frac{m^2}{p_u^2 \tau_3}\right)n^\mu
\delta{\mathscr{M}}_{u\perp,\mu}^{(0)},
\end{eqnarray}
\begin{eqnarray}
\mathscr{S}^{(1)\mu\nu}&=&\frac{1}{p_n} \epsilon^{\mu\nu\alpha\beta}p_\alpha \mathscr{M}_{\beta}^{(1)}
+\frac{1}{2p_n}\left(\Delta^{\mu\lambda} n^\nu - \Delta^{\nu\lambda} n^\mu\right)\partial_\lambda^x {\mathscr{F}}^{(0)}
\nonumber\\
& &+\frac{m}{2p_n p_u}\left(\Delta^{\mu\lambda} n^\nu - \Delta^{\nu\lambda} n^\mu\right)
\sum_s \left(\frac{ u_\lambda}{\tau_{1s}}+\frac{ p_\lambda}{p_u\tau_{2s}}\right)
\delta{\mathscr{J}}_{s,u}^{(0)},\\
{\mathscr{M}}^{(1)\mu}_\parallel &=& \frac{m \bar p^\mu}{\bar p^2}  \sum_s s {\mathscr{J}}_{s,n}^{(1)}.
\end{eqnarray}

These kinetic equations and Wigner functions  can be simplified  if we make the arbitrary
subsidiary  $n^\mu$ and the physical fluid velocity   $u^\mu$ coincide. For brevity, we will replace $u^\mu$ with $n^\mu$ instead of replacing
$n^\mu$  with $u^\mu$. Besides, we also sum  the zeroth and first orders into a unified form by defining
\begin{eqnarray}
\mathscr{J}_{sn} = \mathscr{J}_{sn}^{(0)}+\mathscr{J}_{sn}^{(1)},
 \ \ \ \mathscr{M}^{\mu}=\mathscr{M}^{(0)\mu}+\mathscr{M}^{(1)\mu}.
\end{eqnarray}
With $n^\mu=u^\mu$,  the quantum kinetic equations given above in separate orders  are equivalent to the following form up to the first order $\hbar $,
\begin{eqnarray}
\label{Jsn-eq-0-RTA-2}
p^\mu \partial^x_\mu\left( \frac{\mathscr{J}_{sn}}{p_n}\right)
&=&-\frac{  s}{2p_n}p^\nu \left(\partial^x_\nu  n_\mu \right)
\left[  m \frac{ {\mathscr{M}}^{\mu}}{p_n}
- \frac{1}{p_n}\bar\epsilon^{\mu\alpha\beta} \partial^x_\alpha
\left(\frac{ p_\beta}{p_n} {\mathscr{J}}_{s,n}-\frac{s m}{2p_n}{\mathscr{M}}_\beta\right) \right]\nonumber\\
& &-\left( \frac{1}{\tau_{1s}}+\frac{m^2}{p_n^2\tau_{2s}}\right)
\delta{\mathscr{J}}_{s,n}
+\frac{sm^2}{2p_n^2}\sum_{s'} s' \left( \frac{1}{\tau_{1s'}}+\frac{1}{\tau_{2s'}}\right)
\delta{\mathscr{J}}_{s',n}\nonumber\\
& &-\frac{s}{2p_n^2 \tau_{1s}}\bar\epsilon^{\mu\alpha\beta}p_\beta (\partial^x_\mu n_\alpha)
\delta{\mathscr{J}}_{s,n}\nonumber\\
& &+\frac{m}{4p_n^2}\bar\epsilon^{\mu\alpha\beta}p_\beta \left[ p^\nu (\partial^x_\nu n_\mu)  - p_n  \partial^x_\mu \right]
\left(\frac{\delta{\mathscr{M}}_{\perp\alpha}}{p_n^2\tau_{3}}\right),
\end{eqnarray}
\begin{eqnarray}
\label{M-eq-0-RTA-2}
p^\nu \partial^x_\nu \left(\frac{\mathscr{M}^{\mu}}{p_n}\right)
&=&-\frac{1}{p_n}  p^\lambda   (\partial_{\lambda}^x n_\nu)
\left[ p^\mu \frac{\mathscr{M}^{\nu} }{p_n}
-\frac{m}{2p_n}\bar\epsilon^{\mu\nu\alpha}
\partial_\alpha^x \left( \frac{1}{p_n} \sum_s{\mathscr{J}}_{s,n} \right) \right]\nonumber\\
& &- \left(\frac{1}{\tau_4} + \frac{m^2}{p_n^2 \tau_3}\right)
\delta{\mathscr{M}}^{\mu}_\perp
+\frac{m \bar p^\mu}{p_n^2}\sum_{s'} \frac{s'}{\tau_{1s'}}
\delta{\mathscr{J}}_{s',n}\nonumber\\
& &+\frac{m}{2p_n^2}\bar\epsilon^{\mu\nu\alpha}\left[  p^\lambda   (\partial_{\lambda}^x n_\nu)
-p_n \partial_{\nu}^x\right]
\sum_{s'}\frac{ p_\alpha}{p_n^2\tau_{2s'}}
\delta{\mathscr{J}}_{s',n}.
\end{eqnarray}
The Wigner functions are given by
\begin{eqnarray}
\label{Js-final}
{\mathscr{J}}_{s,\mu}
&=& \frac{ p_\mu}{p_n} {\mathscr{J}}_{s,n}-\frac{s m}{2p_n}{\mathscr{M}}_\mu
+\frac{s}{2p_n}\bar\epsilon_{\mu\alpha\beta} p^\beta \partial_x^{\alpha}\left(\frac{{\mathscr{J}}_{s,n}}{p_n}\right)
+\frac{m}{4p_n^3 \tau_3}\bar\epsilon_{\mu\alpha\beta} p^\beta \delta{\mathscr{M}}^{\alpha}_\perp ,\\
{\mathscr{F}}  &=&  \frac{ m}{p_n} \sum_s{\mathscr{J}}_{s,n}
-\frac{ 1}{2p_n}\epsilon_{\mu\nu\alpha\beta} p^\alpha n^\mu\partial_x^{\nu}\left(\frac{{\mathscr{M}}^{\beta}}{p_n}\right),\\
{\mathscr{P}} &=& -\frac{1}{2p_n}\left(p_\nu n_\sigma - p_n g_{\nu\sigma}\right)
\partial^{\nu}_x\left(\frac{ {\mathscr{M}}^{\sigma}}{p_n}\right)
-\frac{m}{2p_n^2}\sum_s s  \left(\frac{ 1 }{\tau_{1s}}+\frac{1}{\tau_{2s}}\right)
\delta{\mathscr{J}}_{s,n},
\end{eqnarray}
\begin{eqnarray}
\mathscr{S}^{\mu\nu}&=&\frac{1}{p_n} \epsilon^{\mu\nu\alpha\beta}p_\alpha \mathscr{M}_{\beta}
+\frac{m}{2p_n}\left(\Delta^{\mu\lambda} n^\nu - \Delta^{\nu\lambda} n^\mu\right)
\partial_\lambda^x\sum_s \frac{{\mathscr{J}}_{s,n}}{p_n}
\nonumber\\
& &+\frac{m}{2p_n^3}\left(\bar p^\mu n^\nu - \bar p^\nu n^\mu\right)
\sum_s  \frac{1}{\tau_{2s}}\delta{\mathscr{J}}_{s,n},\\
{\mathscr{M}}^{\mu}_\parallel &=& \frac{m \bar p^\mu}{\bar p^2}  \sum_s s {\mathscr{J}}_{s,n}.
\end{eqnarray}
The Wigner functions have been greatly reduced when we identify $n^\mu$ as the local fluid velocity  $u^\mu$. The scalar Wigner function ${\mathscr{F}} $ even has no
explicit dependence on the collision terms. From (\ref{Js-final}) and (\ref{Js}), we can obtain the axial and vector Wigner functions directly
\begin{eqnarray}
\label{A-final}
{\mathscr{A}}_{\mu}
&=& \frac{ p_\mu}{p_n} {\mathscr{A}}_{n}-\frac{m}{p_n}{\mathscr{M}}_\mu
+\frac{1}{2p_n}\bar\epsilon_{\mu\alpha\beta} p^\beta \partial_x^{\alpha}\left(\frac{{\mathscr{V}}_{n}}{p_n}\right),\\
{\mathscr{V}}_{\mu}
&=& \frac{ p_\mu}{p_n} {\mathscr{V}}_{n}
+\frac{1}{2p_n}\bar\epsilon_{\mu\alpha\beta} p^\beta \partial_x^{\alpha}\left(\frac{{\mathscr{A}}_{n}}{p_n}\right)
+\frac{m}{2p_n^3 \tau_3}\bar\epsilon_{\mu\alpha\beta} p^\beta \delta{\mathscr{M}}^{\alpha}_\perp
\end{eqnarray}
which denote the spin polarization distribution and electric charge separation  in phase space, respectively. We notice  that the spin polarization represented by ${\mathscr{A}}_{\mu}$
has no explicit dependence on the collision terms in local comoving frame. The collision dependence has already been totally  coded in the distribution function itself.
The electric charge separation represented by ${\mathscr{V}}_{\mu}$ has dependence on the collisions terms only from the the transverse magnetic momentum distribution $\delta{\mathscr{M}}^{\alpha}_\perp $. This contribution is totally from non-equilibrium effect and could affect chiral magnetic effect when there exists
transverse polarization  in relativistic heavy-ion collisions. Certainly, these conclusions are drawn from relaxation-time approximation and might be changed when we consider more general
collision contribution. However it still shed light on how the collision terms contribute to the spin polarization or electric charge separation in heavy-ion collisions.

In the chiral limit, we can safely set $m=0$ in our formalism of GCKT and obtain the quantum kinetic equations
\begin{eqnarray}
\label{Jsn-eq-0-RTA-2}
p^\mu \partial^x_\mu\left( \frac{\mathscr{J}_{s,n}}{p_n}\right)
&=&\frac{  s}{2p_n^2}p^\nu \left(\partial^x_\nu  n_\mu \right)
\bar\epsilon^{\mu\alpha\beta} p_\beta \partial^x_\alpha \left(\frac{ \mathscr{J}_{s,n}}{p_n}\right)\nonumber\\
& &-\frac{1}{\tau_{1s}}\left[1+\frac{s}{2p_n^2}\bar\epsilon^{\mu\alpha\beta}p_\beta (\partial^x_\mu n_\alpha)\right]
\delta{\mathscr{J}}_{s,n},\\
\label{M-eq-0-RTA-2}
p^\nu \partial^x_\nu \left(\frac{\mathscr{M}^{\mu}_\perp}{p_n}\right)
&=&-\frac{1}{p_n}  p^\lambda   (\partial_{\lambda}^x n_\nu)
p^\mu \frac{\mathscr{M}^{\nu}_\perp }{p_n}
- \frac{1}{\tau_4} \delta{\mathscr{M}}^{\mu}_\perp
\end{eqnarray}
and the Wigner functions read
\begin{eqnarray}
\label{Js-final}
{\mathscr{J}}_{s,\mu}
&=& \frac{ p_\mu}{p_n} {\mathscr{J}}_{s,n}
+\frac{s}{2p_n}\bar\epsilon_{\mu\alpha\beta} p^\beta \partial_x^{\alpha}\left(\frac{{\mathscr{J}}_{s,n}}{p_n}\right),\\
{\mathscr{F}}  &=& -\frac{ 1}{2p_n}\epsilon_{\mu\nu\alpha\beta} p^\alpha n^\mu\partial_x^{\nu}\left(\frac{{\mathscr{M}}^{\beta}}{p_n}\right),\\
{\mathscr{P}} &=& -\frac{1}{2p_n}\left(p_\nu n_\sigma - p_n g_{\nu\sigma}\right)
\partial^{\nu}_x\left(\frac{ {\mathscr{M}}^{\sigma}}{p_n}\right),\\
\mathscr{S}^{\mu\nu}&=&\frac{1}{p_n} \epsilon^{\mu\nu\alpha\beta}p_\alpha \mathscr{M}_{\beta},\\
\mathscr{M}^\mu_\parallel &=& 0
\end{eqnarray}
We see that the Wigner functions $\mathscr{J}_{s,n}$ and $\mathscr{M}^{\mu}$ are completely decoupled with each other  as it should be. It is remarkable that
all the Wigner functions have no explicit dependence on the collision terms at chiral limit.
\section{Summary and discussion}

In this paper, we start from the gauge-invariant  Wigner function formalism in quantum electrodynamics and  derive the quantum kinetic theory
with  BBGKY  hierarchy or general collisions terms. With the help of the semiclassical $\hbar$ expansion, we  use the property of  self-consistency of the Wigner equations to constrain the collision terms up to second order. These constraint equations provide  general
necessary conditions  to make some approximation or simplification in some specific cases.  When only three vector functions involved,   we present a specific solution  for the collision terms in the formalism of GCKT.  We  further specialize this solution in the  form of  relaxation-time approximation and present the self-consistent
quantum kinetic theory at relaxation-time approximation which aligns with the self-consistent constraints.
We find the quantum kinetic theory at relaxation-time approximation can be greatly simplified by defining or decomposing the distribution functions in the comoving fluid frame with velocity $u^\mu$. Some Wigner functions  exhibit even no explicit dependence on the collision terms  and the implicit dependence is totally coded into the distribution functions  when we write the equations or functions in the fluid comoving frame.

Although we didn't take into account the mean electromagnetic field during deriving  the quantum kinetic equation, Wigner functions,
 and the constraint conditions with collision terms,  it is straightforward and  easy to obtain the results when a mean field is
 involved. As we mentioned in the section \ref{sec:wigner}, the collision functions discussed in this work  actually contain all possible terms,
including mean field contribution. At mean field approximation, all the constraint conditions derived in previous sections will be
satisfied automatically as verified in \cite{Gao:2018wmr,Gao:2019znl,Ma:2022ins}. Therefore, we can simply decompose the general electromagnetic field in (\ref{C-full}) into mean field part and quantized field part, separate the mean field part from the quantized field, and directly
 obtain the quantum kinetic equation and Wigner functions with mean electromagnetic field involved. The constraint conditions for collision terms from the quantized field part remain unchanged.

We have presented the quantum kinetic equations and Wigner equations in 8-dimensional phase space, 4-dimensional coordinate space $x^\mu$ plus
4-dimensional momentum space $p^\mu$. In 8-dimensional phase space, the on-shell Dirac delta function is  always involved. We can  integrate
the time-like component of momentum to eliminate the singular  Dirac delta function and obtain the quantum kinetic theory in 7-dimensional
phase space, which can be applied to make numerical calculation directly.
Actually, it is a trivial task to obtain the 7-dimensional  quantum kinetic theory  from the  8-dimensional  one  in sections
 \ref{sec:solution} and \ref{sec:RTA} because  only the onshell Dirac delta functions are involved and there are no derivative terms
 with respect to the momentum when the background electromagnetic field is absent.

 It will be  valuable to solve the self-consistent quantum kinetic equation at relaxation-time
approximation  analytical or numerically. We will postpone these interesting and valuable studies in the future.


\section{Acknowledgments}

This work was supported in part by  the National Natural
Science Foundation of China  under Grant
Nos. 12175123 and 12321005.

\end{document}